\documentclass[twocolumn]{aastex631}

\newcommand\xmm{\textit{XMM-Newton}}
\newcommand\chandra{\textit{Chandra}}

\newcommand\nustar{\textit{NuSTAR}}

\newcommand\xrism{\textit{XRISM}}
\newcommand\nicer{\textit{NICER}}
\newcommand\athena{\textit{Athena}}

\newcommand\pcm{cm$^{-2}$}
\newcommand\pcmq{cm$^{-3}$}

\newcommand\logxi{$\log (\xi$/erg~cm~s$^{-1})$}
\newcommand\lognh{$\log(N_{H}/\rm{cm}^{-2})$}

\shorttitle{Response of the disk wind to X-ray pulsations}
\shortauthors{Kosec et al.}
\graphicspath{{./}{figures/}}

\begin{document}

\title{Constraining the Number Density of the Accretion Disk Wind in Hercules X-1 Using its Ionization Response to X-ray Pulsations}

\correspondingauthor{P. Kosec}
\email{peter.kosec@cfa.harvard.edu}

\author{P. Kosec}
\affiliation{MIT Kavli Institute for Astrophysics and Space Research, Massachusetts Institute of Technology, Cambridge, MA 02139}
\affiliation{Center for Astrophysics | Harvard \& Smithsonian, Cambridge, MA, USA}

\author{D. Rogantini}
\affiliation{Department of Astronomy and Astrophysics, The University of Chicago, Chicago, IL 60637}

\author{E. Kara}
\affiliation{MIT Kavli Institute for Astrophysics and Space Research, Massachusetts Institute of Technology, Cambridge, MA 02139}

\author{C. R. Canizares}
\affiliation{MIT Kavli Institute for Astrophysics and Space Research, Massachusetts Institute of Technology, Cambridge, MA 02139}

\author{A. C. Fabian}
\affiliation{Institute of Astronomy, Madingley Road, CB3 0HA Cambridge, UK}

\author{C. Pinto}
\affiliation{INAF - IASF Palermo, Via U. La Malfa 153, I-90146 Palermo, Italy}

\author{I. Psaradaki}
\affiliation{MIT Kavli Institute for Astrophysics and Space Research, Massachusetts Institute of Technology, Cambridge, MA 02139}

\author{R. Staubert}
\affiliation{Institut für Astronomie und Astrophysik, Universität Tübingen, Sand 1, 72076 Tübingen, Germany}

\author{D. J. Walton}
\affiliation{Centre for Astrophysics Research, University of Hertfordshire, College Lane, Hatfield AL10 9AB, UK}



\begin{abstract}

X-ray binaries are known to launch powerful accretion disk winds that can have significant impact on the binary systems and their surroundings. To quantify the impact and determine the launching mechanisms of these outflows, we need to measure the wind plasma number density, an important ingredient in the theoretical disk wind models. While X-ray spectroscopy is a crucial tool to understanding the wind properties, such as their velocity and ionization, in nearly all cases, we lack the signal-to-noise to constrain the plasma number density, weakening the constraints on outflow location and mass outflow rate. We present a new approach to determine this number density in the X-ray binary Hercules X-1 by measuring the speed of the wind ionization response to time-variable illuminating continuum. Hercules X-1 is powered by a highly magnetized neutron star, pulsating with a period of 1.24~s. We show that the wind number density in Hercules X-1 is sufficiently high to respond to these pulsations by modeling the ionization response with the time-dependent photoionization model \textsc{tpho}. We then perform a pulse-resolved analysis of the best-quality \xmm\ observation of Hercules X-1 and directly detect the wind response, confirming that the wind density is at least $10^{12}$ \pcmq. Finally, we simulate \xrism\ observations of Hercules X-1 and show that they will allow us to accurately measure the number density at different locations within the outflow. With \xrism\ we will rule out $\sim3$ orders of magnitude in density parameter space, constraining the wind mass outflow rate, energetics, and its launching mechanism.

\end{abstract}



\section{Introduction} \label{sec:intro}

Accretion disk winds are wide-angle outflows launched from the disks of accreting compact objects. They may be driven by radiation pressure \citep{Shakura+73, Proga+00}, magnetic fields \citep{Blandford+82}, thermally \citep{Begelman+83}, or by a combination of these different launching mechanisms. As the outflow lifts from the accretion disk, it is illuminated by the hot inner accretion flow and ionized. If it passes across our line of sight towards the ionizing source, the wind can imprint an absorption line spectrum upon the continuum emission from the inner accretion flow, allowing us to detect the presence of the wind and to infer its physical properties. Such absorption spectra can readily be detected via X-rays, an energy band which contains line transitions spanning a very broad range of ionizations and plasma temperatures. 

Ionized outflows were detected to date in most types of accreting systems including active galactic nuclei \citep{Weymann+91, Tombesi+10}, tidal disruption events \citep{Miller+15, Kosec+23b}, X-ray binaries \citep{Neilsen+23}, ultraluminous X-ray sources \citep{Pinto+16, Pinto+23} and even in accreting white dwarfs \citep{Greenstein+82, Prinja+00}.

In X-ray binaries, first hints of disk wind absorption lines were detected in late 90s in the X-ray spectra provided by the CCD detectors on board \textit{ASCA} \citep{Ueda+98, Kotani+00}. These lines were later better resolved by the high-resolution grating instruments onboard the \chandra\ and \xmm\ X-ray telescopes, allowing us to learn more about the outflow properties \citep[e.g.][]{Ueda+04}. Nevertheless, even today, much is still unknown about these phenomena, including their launching mechanisms. More recently, powerful outflows were also detected in X-ray binaries through UV, optical as well as IR observations \citep{Munoz-Darias+19, Sanchez-Sierras+20, Castro-Segura+22}. These newly detected outflows and the X-ray winds may be part of the same phenomenon but showing up at different times due to variations in the ionization field \citep{Munoz-Darias+22}.

The winds seen in X-ray binaries have the potential to carry away large a fraction of the matter originally transferred into the accretion disk by the secondary star \citep[e.g.][]{Lee+02, Kosec+20}. Therefore, they can significantly modify X-ray binary accretion flows \citep{Tetarenko+18, Avakyan+24} and the binary evolution of these systems \citep{Verbunt+93, Ziolkowski+18, Gallegos+23}.

High-resolution X-ray spectroscopy allows us to determine the outflow column density, ionization state and the projected velocity of the wind crossing our line of sight towards the X-ray source. Unfortunately, in a great majority of observations, it does not allow us to measure the wind number density. This is possible only in extraordinary quality X-ray spectra through density-sensitive spectral lines \citep{Miller+06, Tomaru+23}. However, even in these exceptional cases, photo-excitation by UV photons \citep{Mewe+78} must be considered as it can complicate or invalidate the density measurement \citep[e.g., ][]{Jimenez+05, Tomaru+23}.

As a result, it is challenging to determine the distance of the outflow from the compact object, and consequently accurately determine its mass flow rate and energetics. This issue is also challenging the study of the wind launching mechanisms. Wind launching mechanisms predict different launching regions for the outflow (at different distances from the compact object) and thus determining the wind location can hold important clues to its origin.

A promising way to determine the number density is by measuring how the wind responds to changes in the ionizing continuum. The plasma will attempt to adjust its ionization state to a variation in the illuminating radiation in order to return to photoionization equilibrium. The speed of this ionization adjustment directly depends on plasma number density \citep{Krolik+95}. Therefore, the wind density can be measured by tracking the variation of plasma ionization as response to the time variable X-ray flux using a time-dependent photoionization model \citep{Nicastro+99, Garcia+13}. In the last few years, a number of new time-dependent photoionization models were developed to achieve these measurements \citep{Rogantini+22, Luminari+22, Sadaula+23}. Some of these models were recently applied to constrain the number densities of warm absorbers in AGN \citep{Li+23, Gu+23}.

In X-ray binaries, the wind number density is likely much higher than that of warm absorbers in AGN \citep[where densities of $10^{4-7}$ \pcmq\ were measured,][]{Gu+23}. Consequently, its response to any flux variations should be much faster (order of seconds or even shorter), making it challenging to accumulate high-enough data quality X-ray spectra, at high-enough time cadence that would allow us to track the ionization state variations. In other words, an extremely bright X-ray source is required, or an X-ray telescope with an extremely large effective area is needed to track these variations in `real time'.

Here we present a completely different and new approach to this issue. Hercules X-1 (hereafter Her X-1) is an X-ray pulsar with a detected accretion disk wind \citep{Kosec+20}, and a pulsation period of about 1.24~s \citep{Tananbaum+72}. Therefore, the disk wind of Her X-1 is repeatedly illuminated by a periodically time-variable X-ray beam from the pulsar. Assuming its number density is high enough, the wind will adjust its ionization state in response to the time-variable X-ray illumination, during each individual pulsation cycle. Hence, by stacking a large number of pulse periods, and performing a pulse-resolved analysis, we can extract the response of the disk wind to the X-ray pulsation, on the timescale of a single 1.24~s X-ray pulsation without the need for an extremely large effective area telescope.

We further note that since the wind is observed in absorption -- along our line of sight towards the X-ray source, the absorbing plasma sees nearly the same input (time variable) X-ray continuum that we observe, only we observe it modified by the absorber itself. X-rays from other lines of sight (e.g. from large scale X-ray emission) could contribute to our observed X-ray emission without ionizing the wind, but their amount is not likely to be significant compared with primary X-ray emission in the case of Her X-1. Most of this primary emission is compact, originating from much closer to the neutron star than where the wind is located.

\subsection{Hercules X-1}

Her X-1, located at a distance of 6.1 kpc \citep{Leahy+14}, is one of the first known X-ray pulsars, discovered in the early 1970s using the Uhuru satellite \citep{Giacconi+72}. It is most well known for its 35-day cycle of alternating High and Low flux states \citep{Katz+76}, which can be explained by the system being oriented almost edge-on, and exhibiting a warped, precessing accretion disk \citep{Gerend+76, Scott+00}. The precession introduces periods of time during which the outer disk blocks our line of sight towards the inner accretion flow, thus resulting in a Low state, and periods of time during which the neutron star is directly observable, resulting in a High state. Additionally, the warped disk precession results in a time-variable line of sight through the disk wind of Her X-1, uniquely allowing us to directly study its vertical structure and geometry \citep{Kosec+23a}.

The pulsations from Her X-1 have a period of 1.24~s and originate from the accretion column of the neutron star \citep{Davidson+73, Ghosh+79}. The inner accretion disk is disrupted by the magnetosphere of the highly magnetized neutron star \citep[$10^{12}$ G,][]{Truemper+78}, and the infalling matter follows the magnetic field lines, creating the so-called accretion columns above the magnetic poles of the star. The accretion column has a non-spherical radiation beam pattern \citep{Basko+75}, and by rotating alongside with the neutron star, it introduces regular X-ray pulsations at the neutron star rotation period since the magnetic axis is misaligned with the rotation axis.

To judge the potential response of the wind to X-ray pulsation of Her X-1, we first determine the most likely range of wind number density. These limits are estimated from previous time-averaged X-ray observations in Section \ref{sec:density_constraints}. Secondly, in Section \ref{sec:tpho} we use the recently developed time-dependent ionization model \textsc{tpho} \citep{Rogantini+22} to model the expected response of the wind given the obtained density limits. We show that the wind density is high enough to quickly respond to the X-ray pulsations with the period of 1.24~s, at least in some of the archival X-ray observations. We then perform a pulse-resolved analysis of the highest quality X-ray observation of the Her X-1 wind, and directly detect this effect. In Section \ref{sec:data}, we describe our data reduction and preparation methods, and in Section \ref{sec:results} we list our spectral modeling methods and the analysis results.

In 2023, the \xrism\ satellite \citep{XRISM+20} launched and began operations. Thanks to its exceptional spectral resolution in the hard X-ray band ($2-10$ keV), \xrism\ will have excellent capabilities to study the wind response to Her X-1 pulsations. In Section \ref{sec:xrism}, we perform \xrism\ observation simulations, showing that the instrument will easily detect this effect. Finally, in Section \ref{sec:discussion} we discuss these results and in Section \ref{sec:conclusions} we give the conclusions of our study.

\section{Prior constraints on wind density} \label{sec:density_constraints}

In order to model the response of the disk wind in Her X-1 to its X-ray pulsations (in the following Section), we put limits on the wind number density. We derive these limits from other physical wind parameters, which were measured by \citet{Kosec+23a}. In that paper we showed that the wind properties evolve systematically with the precession phase as our line of sight moves due to warped disk precession, and samples the wind at different heights above the disk. In Appendix \ref{app:meanwind}, we estimate representative wind properties at different precession phases, corresponding to representative wind properties at different heights above the warped disk. We estimate representative wind column density, ionization parameter, and the isotropic mass outflow rate. Then we use these `average' parameters to infer lower and upper limits on the wind number density at a range of precession phases, using the following arguments.

The constraint with the highest confidence level (but not the most restrictive) on the lower limit of the wind number density comes from photoionization balance. The spectral modeling allows us to determine the wind column density and its ionization parameter, defined as:

\begin{equation}
    \xi=\frac{L_{\rm{ion}}}{n R^{2}}
\end{equation}

where $L_{\rm{ion}}$ is the ionizing luminosity (measured between 13.6 eV and 13.6 keV), $n$ is plasma density and $R$ the distance of the absorber from the ionizing source. Using the definition of the column density $N_{\rm{H}}=n \Delta R$ where $\Delta R$ is the thickness of the absorber, we can calculate the number density such as:

\begin{equation}
    n=\frac{\xi N_{\rm{H}}^{2}}{L_{\rm{ion}}}  \bigg(\frac{\Delta R}{R} \bigg)^{-2}
    \label{denseq}
\end{equation}

Here the $\Delta R/R$ is the relative thickness of the absorbing layer, which naturally cannot be larger than 1. By taking this limiting value, we can calculate the minimum absorber number density:

\begin{equation}
    n_{\rm{min}}=\frac{\xi N_{\rm{H}}^{2}}{L_{\rm{ion}}}
\end{equation}

\begin{figure}
\includegraphics[width=\columnwidth]{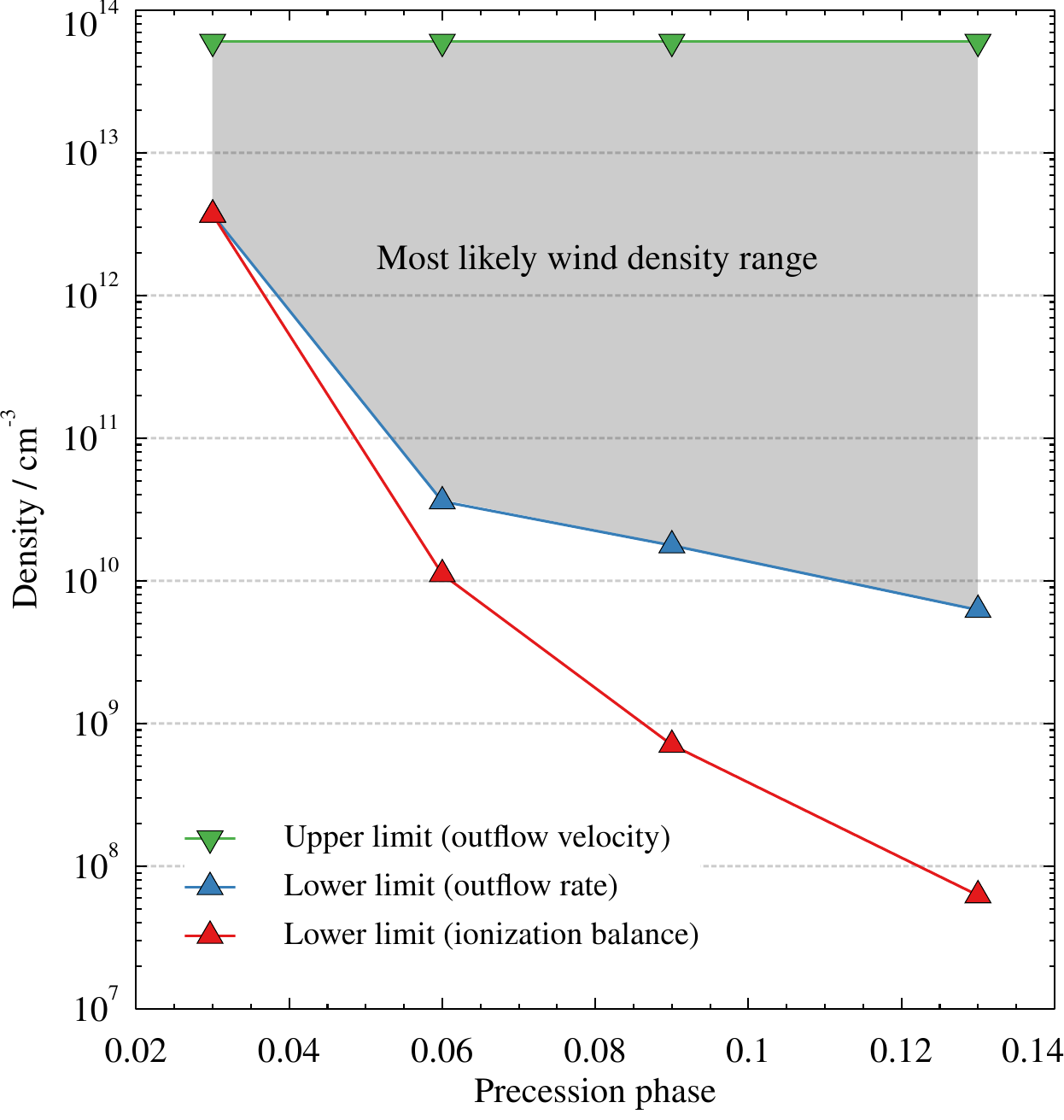}
\caption{The upper and lower limits on the wind number density at different Her X-1 precession phases. The limits are determined by various considerations in Section \ref{sec:density_constraints} using the representative outflow properties measured in the previous studies of Her X-1. The shaded region represents the most likely wind number density range. \label{density_plot}}
\end{figure}

This lower density limit is shown in Fig \ref{density_plot} in red color for a range of disk precession phases.

A stricter lower limit on the wind number density is obtained by considering the maximum mass outflow rate. First, the outflow rate cannot exceed the mass transfer rate through the outer accretion disk (assuming a long-term stable accretion flow). Secondly, the mass outflow rate at greater heights above the disk (further along the streamlines) should not exceed the outflow rate at lower heights. Following the derivation of \citet{Kosec+20}, the mass outflow rate can be calculated from the measured wind quantities as:

\begin{equation}
\label{eqfinalMout}
\dot{M}_{\rm{out}} = 4 \pi \mu m_{\rm{p}} v \frac{L_{\rm{ion}}}{\xi}  C_{\rm{V}}  \frac{\Omega}{4 \pi} = \dot{M}_{\rm{iso}}  C_{\rm{V}}  \frac{\Omega}{4 \pi}
\end{equation}

where $v$ is the wind velocity, $C_{\rm{V}}$ is the volume filling factor and $\Omega$ is the solid angle into which the outflow is being launched, $\mu$ defines the mean atomic mass ($\sim1.2$ assuming solar abundances) and $m_{\rm{p}}$ is the proton mass. $\dot{M}_{\rm{iso}}$ is the isotropic mass outflow rate assuming a maximum volume filling factor (100\%) and a $4\pi$ launching solid angle. \citet{Kosec+23a} found that the volume filling factor $C_{\rm{V}}$ must decrease with precession phase. Otherwise, the mass outflow rates at greater heights above the disk would by far exceed the mass transfer rates through the disk itself as well as the outflow rates at lower heights.

Therefore, by assuming that the solid angle of the outflow does not change with height above the disk, and the fact that the mass outflow rate should not increase upwards in the flow, we can calculate the ratio of volume filling factors $C_{\rm{V}}$ for points at various heights above the disk. We use the calculated $\dot{M}_{\rm{iso}}$ rates (Appendix \ref{app:meanwind}) to get the ratios of $C_{\rm{V}}$, which should be proportional to the relative thickness of the outflow $\Delta R/R$ (since $\Omega\sim$constant). As a result, we can determine the relative thickness ratio for wind measurement points in comparison with the point at precession phase 0.03 (see Fig. \ref{density_plot}), which can be substituted in Eq. \ref{denseq}. This allows us to tighten the photoionization balance limits on the wind density, by taking the most relaxed case scenario of $\Delta R/R=1$ at precession phase of 0.03. These tightened density lower limits are shown in Fig \ref{density_plot} in blue color. However, we note that this is not a hard lower limit on the wind density (in contrast to the photoionization balance limit). It is possible to avoid this limit if the wind solid angle $\Omega$ varies with height above the disk.

Finally, a soft upper limit on the number density can be derived from the outflow velocity and by assuming that this is not a failed outflow. The projected line-of-sight velocity of the Her X-1 wind is mostly between 200 and 800 km/s. Following the calculations of \citet{Kosec+23a}, even assuming the maximum relative thickness ($\Delta R/R=1$), which results in the maximum possible outflow distance from the neutron star, the escape velocity near the base of the wind (at low heights above the disk) is about 2000 km/s. The outflow can still be successfully launched, as we are only observing the line-of-sight velocity component, and do not measure any vertical (due to the high inclination of Her X-1) and especially toroidal (Keplerian) components. However, it is unlikely that the outflow of such a low projected velocity could be launched at much lower radii in the disk, where the required escape velocity is much higher - the velocity projection angle would have to be extreme. 

For this reason, we can put a soft lower limit on the outflow location (which in turn corresponds to a soft upper limit on number density) by considering the outflow at the point where it is observed closest to the neutron star (at the lowest precession phase, 0.03 in Fig \ref{density_plot}). Here we assume that the outflow can only be launched from a location where the escape velocity is less than 4000 km/s, double the velocity that we obtain by taking the maximum outflow distance at these low precession phases. If the outflow is located even closer to the neutron star than this assumption and it is not a failed wind (i.e. its velocity is above 4000 km/s) and it shows a typical line-of-sight velocity of 500 km/s, the wind streamline direction must make an angle of at least 83$^{\circ}$ with respect to our line of sight.

From this assumption of maximum escape velocity we can calculate the ratio of the maximum and the minimum outflow distance, which is 4. This is also the inverse of the minimum outflow relative thickness: $\Delta R/R=0.25$. Finally, using Equation \ref{denseq}, we can calculate the wind density at this minimum distance point, which corresponds to the maximum possible wind density of about $6\times10^{13}$ \pcmq\ at the precession phase of 0.03. At greater precession phases than 0.03, corresponding to greater heights above the disk, it is unlikely that the plasma density would increase compared with the density at the lower heights. Therefore, $6\times10^{13}$~\pcmq\ is the upper limit on the wind density throughout all precession phases. It is shown in green in Fig. \ref{density_plot}. We stress that this is a very soft upper limit on the outflow number density. Its value could in practice be higher. However, in that case the wind would most likely be a failed outflow, and not escape the system at all. This was suggested by \citet{Nixon+20}.

We conclude that the most likely wind density is between $10^{9}$ and $10^{14}$ \pcmq\ across all heights above the disk. The density must be rather high (about $10^{13}$ \pcmq) near the base of the wind, but it may decrease significantly as the outflow climbs to greater heights above the disk. Such high number densities will result in much faster ionization responses (timescales of seconds, similar to the Her X-1 pulse period, or even lower) compared with warm absorbers in AGN, where densities of $10^{4-7}$ \pcmq\ result in the response times of $10^{3-5}$ s or longer \citep{Gu+23, Li+23}. 

\section{Time-dependent photoionization modeling} \label{sec:tpho}
 
In the previous section, we estimated the most likely wind number density range at different phases of the disk precession cycle. Now we can simulate how the wind will respond to X-ray flux variations at the 1.24~s pulsation period. We use the recently developed time-dependent photoionization model \textsc{tpho} \citep{Rogantini+22}, which is publicly available in the \textsc{spex} package \citep{Kaastra+96}. \textsc{tpho} simulates the response of a slab of plasma, which is originally assumed to be in photoionization balance, to a time variable lightcurve. If provided with a constant lightcurve, \textsc{tpho} will reproduce an equivalent transmission spectrum to the one calculated by the equilibrium photoionization spectral model \textsc{pion} \citep{Mehdipour+16} in \textsc{spex}.

We simulate the response of the disk wind to X-ray pulsation at Her X-1 precession phase of about 0.03, when the wind column density is roughly $10^{23}$ \pcm\ and its ionization parameter is \logxi=3.77 (Appendix \ref{app:meanwind}). This precession phase is chosen because the wind column density is at its highest, and so the wind absorption lines are the strongest (and most easily detectable). 

Another input to \textsc{tpho} is the illuminating X-ray continuum. We take the best-fitting continuum spectral model of observation 0865440101 (the highest quality \xmm\ observation of Her X-1 disk wind), which we analyzed in \citet{Kosec+23a} and which occurred exactly at precession phase of 0.03. 

The X-ray pulsation lightcurve is obtained by performing a pulse-resolved analysis of the EPIC pn\footnote{EPIC pn is the best instrument for this analysis with sufficient temporal resolution to resolve the 1.24~s pulsation.} data from this observation, described in the following two sections of this paper. We take the unabsorbed $0.5-10$ keV luminosity of Her X-1 after accounting for disk wind and Galactic absorption (see Section 5 for further details) as the input lightcurve for \textsc{tpho}. We do not use the X-ray count rate as this quantity is affected by the energy-dependent effective area of \xmm, considering the variations in the spectral shape of Her X-1 continuum over the pulsation period. We also do not use the extrapolated total ($0.1-100$ keV) Her X-1 luminosity because this is not as accurately measured by EPIC pn as the narrower $0.5-10$ keV range (which is fully covered by EPIC pn). Using the $0.5-10$ keV band is a conservative estimate of the level of flux variations in Her X-1 considering that its pulsed fraction increases with energy \citep{Ferrigno+23}. Including harder energies would thus increase the flux variation (the total pulsed fraction). Larger flux variations lead to greater changes in wind ionization and so our approach leads to a conservative estimate of how strongly the wind ionization can vary over the Her X-1 pulse period.

The final input of \textsc{tpho} is the plasma number density, which determines how quickly the outflow can react to any variation of ionizing luminosity. We test the wind response for a range of wind densities based on the findings of the previous section: $10^{9}$, $10^{10}$, $10^{11}$, $10^{12}$, $10^{13}$, $10^{14}$ and $10^{15}$ \pcmq. Such a broad range of values should encompass any possible Her X-1 wind density, even accounting for uncertainties in the approach adopted in the previous section.

One significant drawback of the current version of \textsc{tpho} is that the model assumes a constant spectral energy distribution (SED) of the illuminating continuum throughout the time evolution of plasma ionization state. This approximation is not fully valid for Her X-1, where the X-ray spectrum varies with pulse phase. Specifically, the soft X-rays pulsate out of phase in comparison with hard X-rays \citep[e.g.][]{Zane+04}. This spectral variation is shown in Appendix \ref{app:PNspec}. Nevertheless, \textsc{tpho} will still provide an extremely useful view on the wind plasma response to Her X-1 X-ray pulsation. However, given the SED limitation at this stage, we prefer not to directly fit X-ray spectra with \textsc{tpho} to determine the best-fitting wind density, as this measurement would not be reliable. A future version of \textsc{tpho} (Rogantini et al. in prep) will include a time-variable source SED. 

We input the spectral continuum model, the disk wind properties, its number density and the X-ray pulsation lightcurve to \textsc{tpho}. Then the plasma is evolved for 5 X-ray pulsation cycles (about 6~s in real Her X-1 pulsation time) to obtain a stable wind response. Afterwards, the wind properties are extracted from the following two X-ray pulsation cycles. We particularly follow the Fe XXV and Fe XXVI ion column densities, as the Fe XXV (6.67 keV) and Fe XXVI (6.96 keV) transitions are the strongest resolved disk wind absorption lines in the EPIC pn energy band. We note that the output from \textsc{tpho} simulations are the ionic concentrations of different ions relative to hydrogen. We convert these concentrations to ionic column densities using the best-fitting wind column density ($N_{H}=10^{23}$ \pcm) and the best-fitting Fe abundance from observation 0865440101. This Fe abundance ($\rm{Fe/O}=2.1$) was determined by a time-averaged wind abundance analysis by \citet{Kosec+23a}. This simulation procedure is performed for all the wind number densities tested.

\begin{figure}
\includegraphics[width=\columnwidth]{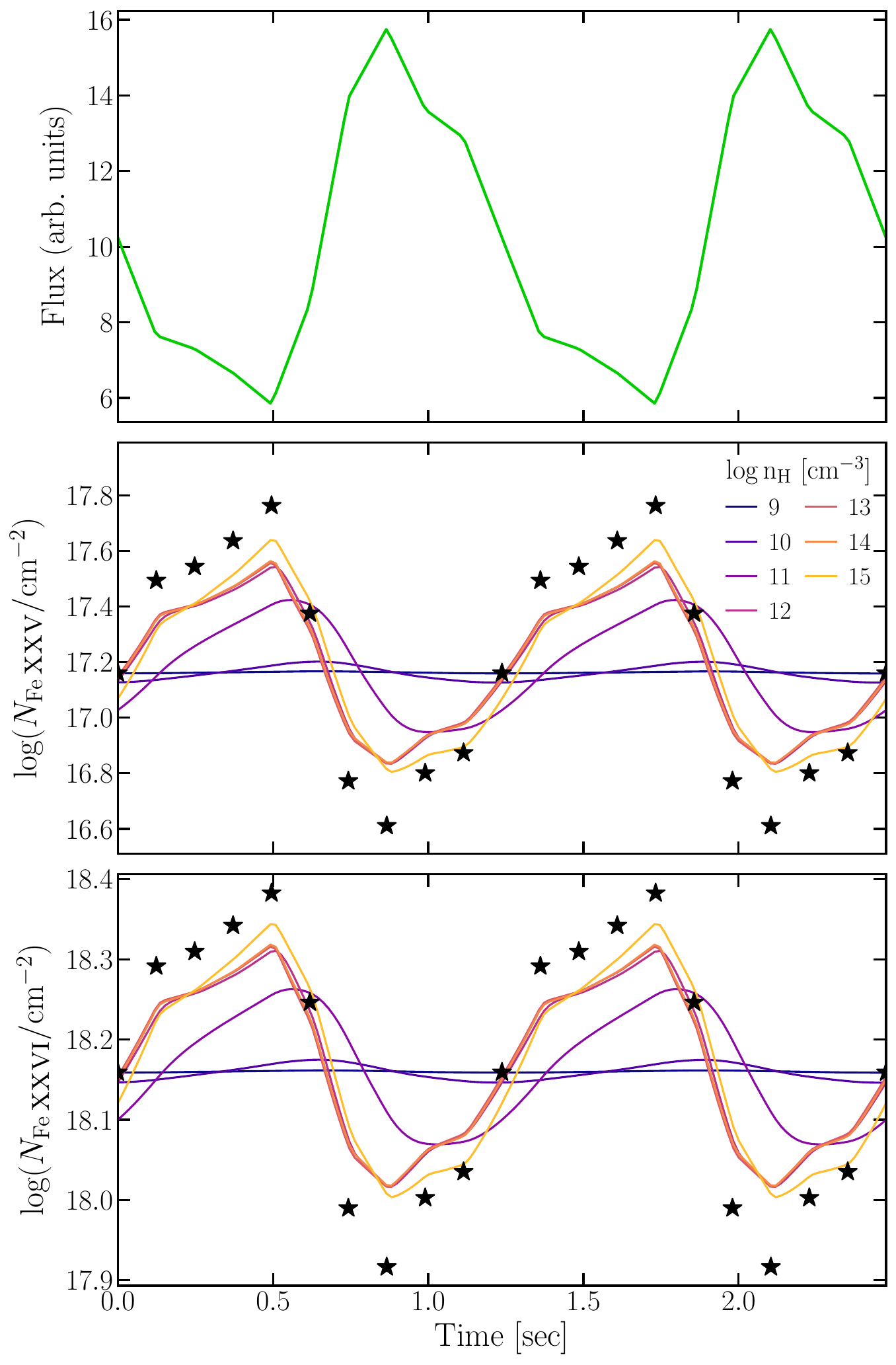}
\caption{\textsc{tpho} simulations of ionized disk wind response to Her X-1 X-ray pulsations. The top panel shows the intrinsic $0.5-10$ keV pulsation lightcurve (2 cycles are shown). The middle and the bottom panels show the column densities of Fe XXV and Fe XXVI ions, respectively. The black stars show the response of plasma in immediate photoionization equilibrium, while the lines of different colours show plasma of different number densities (from $10^{9}$ to $10^{15}$ \pcmq). \label{tpho_results}}
\end{figure}

We show the simulation results in Fig \ref{tpho_results}. The top panel shows the input lightcurve (of unabsorbed $0.5-10$ keV luminosity) over the X-ray pulsation period (two cycles are shown). The middle and the bottom panels show the column density in the Fe XXV and Fe XXVI ions, respectively, over the pulse period. The colours show the response of plasma of different number densities, while the stars show the column densities for plasma in immediate photoionization equilibrium. For low plasma densities ($10^{10}$ \pcmq\ and lower), the wind density is so low that it basically does not respond to the X-ray pulsation at all. For higher, but still low densities ($10^{11}$ \pcmq) the response is weak and significantly delayed in comparison with the pulsation lightcurve. For higher densities ($10^{12}$ \pcmq\ and higher), the response becomes faster and stronger, more in line with the response of plasma in immediate photoionization balance.

The shape of the Fe XXV and XXVI responses can be explained by analyzing how their ionic concentrations respond to a change in the plasma ionization parameter. This is explored in Appendix \ref{app:ion_con}. From Fig. \ref{ion_con}, we can see that in this high ionization regime at \logxi\ of around 3.8, most Fe atoms are fully ionized (in the Fe XXVII state). Assuming the number density is sufficient, an increase in incident flux leads to a rise in the ionization parameter. This results in a decrease in the ionic concentrations of both Fe XXV and XXVI, reducing their column densities. A return into original (lower) incident flux decreases the ionization parameter. As a result, some of the Fe XXVII atoms recombine into the Fe XXV and XXVI states, increasing the column densities in these transitions.

\begin{figure}
\includegraphics[width=\columnwidth]{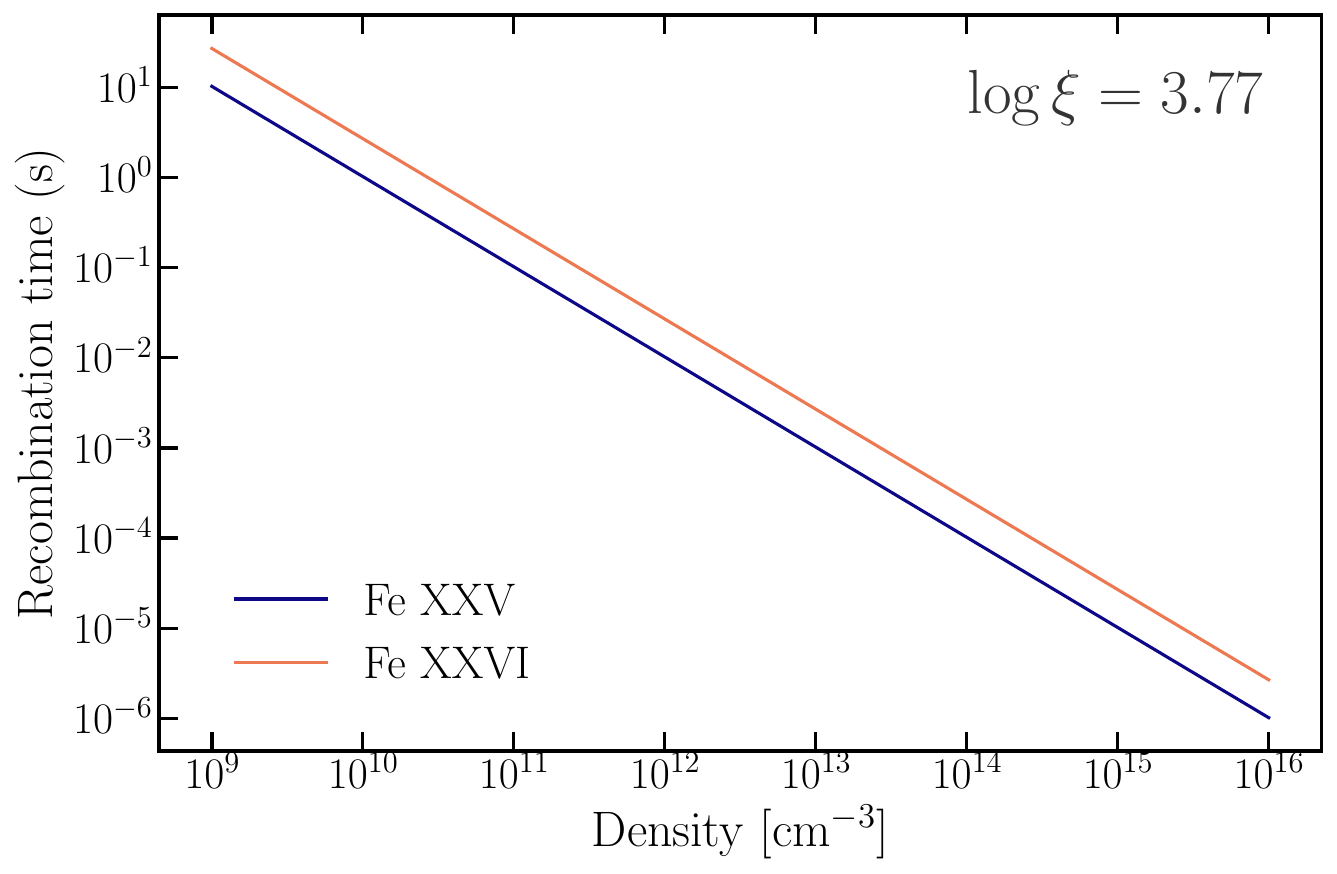}
\caption{Fe XXV (blue) and XXVI (orange) ion recombination times versus plasma number density $n_{\rm{e}}$, which can be calculated as $t_{\rm{rec}} = (\alpha_{\rm{rec}} n_{\rm{e}})^{-1}$, where $\alpha_{\rm{rec}}$ is the radiative recombination coefficient \citep[e.g.][]{Nicastro+99}.  \label{tpho_trec}}
\end{figure}

An alternative way to show the speed of disk wind response to any variations in ionizing flux is through the recombination times in different transitions. This is shown in Fig \ref{tpho_trec} for the Fe XXV and Fe XXVI transitions, assuming wind properties similar to those at precession phase of 0.03. For low wind densities ($n<10^{10}$ \pcmq), the wind response is longer than 1~s, as long or longer than the 1.24~s pulsation period. For high number densities ($10^{12-13}$ \pcmq\ and higher), both the Fe XXV and XXVI recombination times are very fast, lower than 0.01~s, and it will be challenging to measure them directly.

Given the wind density limits we obtained in the previous section, we can make the following predictions about the wind response to X-ray pulsation of Her X-1. At low precession phases, the density is expected to be high (above $10^{12}$ \pcmq) and so the wind should respond to the pulsation and it should respond almost immediately. The strength of the response is related to the ratio of the minimum and the maximum illuminating flux during the pulsation cycle. For plasma in immediate photoionization equilibrium, a ratio of $F_{max}/F_{min}\sim2-3$ (appropriate for Her X-1), should correspond to a change in the ionization parameter $\Delta$\logxi$\sim0.3-0.5$. This effect should therefore be directly observable, even with current X-ray telescopes, as will be shown in the following sections.

At higher precession phases, the wind absorption is weaker \citep{Kosec+23a}, complicating the possibility of detection of any pulse-resolved variability. However, the wind density may also be lower, as low as $10^9$ \pcmq\ for precession phases around 0.13. If the wind density indeed decreases with height above the disk, as is likely, its density may cross from the range when the response is immediate, through a range when the response is significantly delayed and washed out, to a range when it stops responding altogether to the X-ray variations.

\section{Data Reduction and preparation} \label{sec:data}

The wind absorption lines and their strengths evolve with the Her X-1 precession phase. As the measurement of the variation in the wind ionization state over the pulsar rotation period is quite challenging using the current X-ray instruments, we analyse only the observation with the highest signal-to-noise in the wind absorption lines. This is \xmm\ \citep{Jansen+01} observation 0865440101, which occurred very close after the Her X-1 Turn-on into the Main High state, and covers low precession phases around $0.02-0.04$. The observation began on August 10, 2020 (MJD=59071) and lasted for 126ks. The Reflection Grating Spectrometer \citep[RGS,][]{denHerder+01} lightcurve of this observation is shown in the left part of Fig. 2 in \citet{Kosec+22}. For this analysis we only extract the high-flux period with a duration of about 50 ks in the second half of the observation, not including a dipping period at the end of the observation. We do not analyze the low-flux periods as their signal-to-noise is much lower. Additionally, the dipping periods contain highly variable, low-ionization and neutral absorption originating at the outer edge of the Her X-1 disk, complicating the spectral analysis of the disk wind, and so are also ignored.

The observation files were downloaded from the XMM-Newton Science Archive and reduced using \textsc{sas} V20, \textsc{caldb} as of February 2023. We perform a pulse-resolved extraction during which we split the X-ray period of Her X-1 into 10 pulse phase bins of equal length (about 0.124~s). The number of phase bins was chosen as a compromise between the bin time resolution and the signal-to-noise. In order to perform this split accurately, we first determined the precise Her X-1 pulsation period at the time of the observation. Initially, the photon arrival times were corrected into solar system barycenter using the \textsc{barycen} tool with the DE200 ephemeris. Then we corrected the photon arrival times for the orbital motion of Her X-1 using the orbital solution of \citet{Staubert+09}. Finally, we used the \textsc{epoch\_folding\_search} function within the Stingray package \citep{Huppenkothen+19, Stingray+19} to determine the precise period of Her X-1 during observation 0865440101. By locating the peak of the $\chi^2$ epoch folding (EF) statistics, we found P=1.23771944(2)~s, which is in very good agreement with the measurements around MJD=59071 (10 August 2020) by the Fermi/GBM Pulsar Project \citep{Finger+09}. Using this timing solution, we produced Good Time Interval (GTI) files containing the split times for the 10 pulse phase bins throughout the full \xmm\ observation.

We use data from the European Photon Imaging Camera (EPIC) pn instrument \citep{Struder+01}. We are unable to use any RGS data for the pulse-resolved analysis because the frame time of RGS, which was operated in the standard spectroscopy mode is too long (frame time of 4.8~s for RGS1 and 9.6~s for RGS 2). We also do not use EPIC MOS data as they are too piled up. EPIC pn was used in the Timing mode during the analysed observation, to minimize the effect of pile-up. The data were reduced using the \textsc{epproc} routine and only events of PATTERN$\leq$4 (single and double) were accepted as valid events. We produced a background lightcurve of the analysed observation segment using the \textsc{evselect} routine but did not identify any periods of high background flaring.

Despite the usage of Timing mode, some of the EPIC pn spectra are piled-up. The pulse-resolved analysis allows us to apply different source regions for different pulse phase bins depending on their photon count rate (and the amount of pile-up). For those phase bins not affected significantly by pile-up, the source regions are rectangles centered on the core of the point-spread function (PSF). For the two phase bins significantly affected by pile-up (bins 6 and 7, with uncorrected count rates of $700-800$ s$^{-1}$), we removed the central two columns. In both cases, the background regions were rectangular regions composed of a few columns as far away from the core of the source PSF as possible. Since Her X-1 is an extremely bright source, the background flux was not comparable to the source flux at any energies over the bandpass used here for any of the pulse phase bins. All phase bins have net count rates (after any pile-up correction) of $380-480$ counts/s, and each bin has a net exposure of 4250~s. Further details about the individual phase bins are given in Table \ref{pndata_table}.

\begin{deluxetable}{ccc}
\tablecaption{Details of pulse-resolved data extraction. For each pulse phase bin, the net count rates are listed after any pile-up correction as well as the number of central columns excluded from the source region to mitigate pile-up.\label{pndata_table}}
\tablewidth{0pt}
\tablehead{
\colhead{Pulse phase bin} & \colhead{Count Rate / s} & \colhead{Columns Excluded}  
}
\startdata 
0&402.9 & 0\\
1&390.5 & 0\\
2&438.7 & 0\\
3&438.5 & 0\\
4& 389.7& 0\\
5& 468.7& 0\\
6& 388.8&2 \\
7&400.3 &2\\
8&481.6 & 0\\
9& 451.5& 0\\
\enddata
\end{deluxetable}

The reduced spectra were binned to oversample the instrumental EPIC pn spectral resolution by at most a factor of three, and to at least 25 counts per bin using the \textsc{specgroup} routine. The full energy range used is 0.5 to 10 keV. At these high photon fluxes, EPIC pn suffers from a gain shift issue, which is described specifically for Her X-1 in Appendix A of \citet{Kosec+22}, and more generally by \citet{Duro+16}. Our mitigation strategy for this gain shift follows our approach in \citet{Kosec+22} and is also described in Section \ref{sec:results}. 

Finally, all the reduced spectra were converted from \textsc{ogip} format into \textsc{spex} format using the \textsc{trafo} routine. We used \textsc{spex} \citep{Kaastra+96} v3.07 to fit X-ray spectra and used the Cash statistic \citep{Cash+79} for analysis. Use of the Cash statistic is not formally necessary for any of our EPIC-pn data analysis (where the number of counts per bin is very high), but it is appropriate for \xrism\ data simulation analysis. All uncertainties are quoted at $1\sigma$ significance.

\section{Spectral Modeling and Results} \label{sec:results}

After the pulse-resolved spectra are prepared, we proceed with spectral modeling to determine how the disk wind properties vary over the pulsation cycle. Her X-1 shows strong variability in the $0.5-10$ keV EPIC pn energy band over the X-ray pulsation period, both in the overall count rate and in the spectral shape. We show the spectra of a number of pulse phase bins in Fig. \ref{PNspecfig} in Appendix \ref{app:PNspec}. In particular, we notice that the hard X-rays (primarily direct emission) pulsate with a phase shift compared with the soft X-rays (primarily reprocesing of direct emission). This effect is well known and was previously studied by \citet{Zane+04} and \citet{brumback+21}.

\subsection{Spectral model setup}

We use a simplified spectral model from \citet{Kosec+22} to fit the Her X-1 continuum and emission components in the EPIC pn energy range. Initially, we use the full $0.5-10$ keV range of EPIC pn to constrain the Her X-1 spectral shape. A \textsc{comt} component \citep{Titarchuk+94} is used to describe the hard Comptonization continuum originating from the accretion column of the neutron star. A soft blackbody with a temperature of about 0.1 keV describes the thermal reprocessing signature \citep{Hickox+04} observed at soft X-rays ($<1$ keV). Additionally, we use a Gaussian line to describe the broad `1 keV' feature \citep[e.g.][]{Fuerst+13, Kosec+22} in the soft band. The Fe K band around 6.4 keV is very complex in Her X-1 - following the results of \citet{Kosec+22}, we describe it using 3 Gaussian lines of different line widths: a narrow Gaussian (low ionization Fe emission) with a Full Width Half Maximum (FWHM) of 0.05 keV (fixed) and an energy of 6.4 keV (fixed); a medium-width Gaussian (Fe XXV emission) with a width of about 0.5 keV (free) and energy of 6.67 keV (fixed); and a broad Gaussian line (Fe K emission, unknown ionization state) with an energy of about 6.5 keV (free), and a width of about 2 keV (free).

All of the emission components are absorbed by the ionized disk wind plasma, the modeling of which is discussed further below. On top of this absorption, we apply Galactic absorption using the \textsc{hot} model, with the absorption column density fixed to $1\times10^{20}$ \pcm\ \citep{Kosec+22}. This model describes an almost neutral absorber with a temperature of $8\times10^{-6}$ keV \citep{dePlaa+04, Steenbrugge+05}. 

Finally, to correct for the gain shift issue in EPIC pn mentioned in section \ref{sec:data}, on top of all the previous spectral models, we apply a multiplicative shift model \textsc{reds} which blueshifts all the spectral models by a certain factor (a free variable of the spectral fit), and thus introduces a variable gain shift to the spectral model, to match with that of the EPIC pn data. Further details about this gain shift correction can be found in Appendix A of \citet{Kosec+22}. The value of the gain shift parameter is anchored in the EPIC pn data by the fixed energies of certain Fe emission lines (low ionization Fe and Fe XXV) and by the positions of the disk wind absorption lines (Fe XXV and Fe XXVI). The projected velocity of the wind ($600 \pm 40$ km/s) was accurately measured in previous time-averaged analysis of this observation using both EPIC pn and RGS data \citep{Kosec+23a} and so the energies of the Fe XXV and XXVI lines are known. Outside of the Fe K region, this gain correction may not be completely accurate since applying a simple blueshift parameter may not completely describe the true EPIC pn gain shift function. However, outside of this region, our spectral model in the $2-10$ keV band is primarily just a simple, featureless powerlaw shape which does not require a perfect understanding of the gain function.

The disk wind produces strong absorption lines in both the Fe K band (Fe XXV and Fe XXVI) and in the soft X-ray band (N VII, O VIII, Ne X). The soft X-ray absorption is well resolved in the RGS spectra \citep{Kosec+20, Kosec+23a}, but it is completely unresolved in EPIC pn which has a spectral resolution of $\sim100$ eV throughout the $0.5-10$ keV bandpass. At the same time, the soft X-ray band is very complex in Her X-1, containing an array of both narrow and broad emission lines in addition to wind absorption \citep{Kosec+22}. For these reasons, we only use the soft X-ray ($<2$ keV) EPIC pn data to initially describe the broadband continuum, and constrain the X-ray spectral energy distribution (SED) of Her X-1. An accurate description of the SED is necessary for photoionization modeling of the disk wind. After the soft X-ray shape is reasonably constrained, we ignore the EPIC pn data below 2 keV, fix any spectral parameters relevant to this band (the soft blackbody, 1 keV line properties and the seed temperature of the Comptonization component) and re-fit the resulting $2-10$ keV spectra. With this approach, any unresolved features below 2 keV (disk wind absorption and any emission lines) will not sway our spectral fit of disk wind absorption in the crucial Fe K band.

We describe the wind absorption features in the Fe K band and study the response of the wind to X-ray pulsation using two methods. First with a phenomenological approach where we measure the column density in the Fe XXV and XXVI ions, and second with a physical photoionization model, which allows us to measure the plasma ionization parameter \logxi. We will compare whether these two methods yield consistent results.

Before proceeding to spectral fitting, we consider the results of the time-averaged analysis of the full observation 865440101 (the same observation as used in this work), which was performed by \citet{Kosec+23a}. In that work, the RGS and EPIC pn data from this observation were fitted (without performing any pulse-resolved extraction) with the photoionization model \textsc{pion} (see further details about this model below) and the full continuum model from \citet{Kosec+22}. This is the highest-fidelity measurement of the parameters of the disk wind in Her X-1, thanks to the excellent signal-to-noise (S/N) of this (full) observation and the simultaneous usage of both RGS and EPIC pn data. The pulse-resolved analysis is likely going to be more limited, as the S/N of the original observation is split into 10 pulse phase bins, and importantly we are unable to use the simultaneous RGS spectra since the frame time of the RGS instrument (in standard spectroscopy mode) is longer than the pulse period of Her X-1.

The time-averaged analysis allowed \citet{Kosec+23a} to determine the following wind photoionization properties: a column density of  $1.00^{+0.09}_{-0.05} \times 10^{23}$ \pcm, an ionization parameter \logxi\ of $3.77_{-0.02}^{+0.03}$, a velocity width of $186 \pm 15$ km/s, and a systematic velocity of $-600 \pm 40$ km/s (here given uncorrected for Her X-1 orbital motion). At such high ionization parameters the huge majority of the Fe K wind absorption will be in the Fe XXV and Fe XXVI transitions, and in the $2-10$ keV energy band the ratio of these two transitions will be the driver for the ionization parameter measurement. 


Finally, in order to compare the pulse-resolved disk wind properties with the X-ray pulsations of Her X-1, as well as for \textsc{tpho} simulations (Section \ref{sec:tpho}), we calculated the $0.5-10$ keV unabsorbed luminosity for each pulse phase bin during the physical photoionization modeling analysis, using the best-fitting X-ray continuum. 


\subsection{Phenomenological disk wind modeling}

The first pulse-resolved modeling approach is phenomenological and we apply the \textsc{slab} model \citep{Kaastra+02} within \textsc{spex} to describe the disk wind features. \textsc{slab} calculates the transmission through a slab of plasma, where all ionic column densities are allowed to vary independently, and are only tied by a common systematic velocity, line velocity width and the covering fraction. This model gives us an opportunity to measure column densities only in certain ions, which are less prone to full energy band spectral behaviour (e.g. SED variations) and therefore systematic uncertainties in the more complex photoionization models. We use \textsc{slab} to determine the wind column density in the Fe XXV and Fe XXVI ions for each pulse phase bin, in order to probe how they respond to changes in the illuminating X-ray continuum. We fix the outflow velocity and the velocity width of the component to the values obtained from the time-averaged photoionization analysis. These quantities are very unlikely to vary systematically over the (very short) pulse period because the orbital period of particles within the outflow is not related to the neutron star rotation period (the outflow is located far beyond the neutron star magnetosphere).

\begin{deluxetable*}{cccccccc}
\tablecaption{Best-fitting disk wind properties from the phenomenological and physical analysis of EPIC pn data from observation 0865440101. The $0.5-10$ keV luminosity is calculated from the observed X-ray flux after accounting for disk wind and Galactic absorption.  \label{pnresults_table}}
\tablewidth{0pt}
\tablehead{
\colhead{Pulse} & \colhead{Time} & \colhead{$0.5-10$ keV} & \multicolumn2c{\textsc{slab} model} & \multicolumn3c{\textsc{pion} model}  \\
\colhead{Bin} & \colhead{} & \colhead{luminosity} & \colhead{log Fe XXV} & \colhead{log Fe XXVI} & \colhead{\logxi}   &\colhead{log Fe XXV} & \colhead{log Fe XXVI}  \\
\colhead{} & \colhead{} & \colhead{} & \colhead{column density$^{1}$} & \colhead{column density$^{1}$} & \colhead{}   &\colhead{column density} & \colhead{column density{}}  \\
\colhead{} & \colhead{s} & \colhead{$10^{36}$ erg~s$^{-1}$} & \colhead{\pcm} & \colhead{\pcm} & \colhead{}   &\colhead{\pcm} &  \colhead{\pcm} 
}
\startdata 
0 & 0 & $ 10.2^{+0.09}_{-0.10} $ &  $ 17.36^{+0.44}_{-0.47} $ & $ 18.51^{+0.14}_{-0.12} $ & $ 3.55^{+0.10}_{-0.09} $    & $ 17.43^{+0.21}_{-0.24} $ &  $ 18.29^{+0.08}_{-0.10} $   \\
1 & 0.124& $ 7.65^{+0.14}_{-0.14} $ &  $ 17.54^{+0.33}_{-0.35} $ & $ 18.27^{+0.19}_{-0.23} $ & $ 3.56^{+0.12}_{-0.10} $ & $ 17.30^{+0.23}_{-0.32} $ &  $ 18.23^{+0.09}_{-0.14} $   \\
2 & 0.248& $ 7.29^{+0.28}_{-0.41} $ &  $ 16.95^{+0.52}_{-3.95} $ & $ 18.22^{+0.20}_{-0.20} $ & $ 3.68^{+0.14}_{-0.12} $ & $ 16.94^{+0.33}_{-0.41} $ &  $ 18.08^{+0.15}_{-0.19} $   \\
3 & 0.371& $ 6.65^{+0.56}_{-0.72} $ &  $ 16.1^{+1.2}_{-3.1} $ & $ 18.11^{+0.17}_{-0.19} $ & $ 3.64^{+0.21}_{-0.17} $ 	& $ 16.85^{+0.46}_{-0.60} $ &  $ 18.04^{+0.20}_{-0.29} $   \\
4 & 0.495& $ 5.83^{+0.54}_{-0.52} $ &  $ <16.75 $ & $ 18.07^{+0.14}_{-0.14} $ & $ 3.58^{+0.15}_{-0.15} $ 	& $ 16.86^{+0.44}_{-0.42} $ &  $ 18.03^{+0.19}_{-0.20} $   \\
5 & 0.619& $ 8.51^{+0.09}_{-0.09} $ &  $ 16.53^{+0.80}_{-3.54} $ & $ 18.25^{+0.21}_{-0.21} $ & $ 3.98^{+0.15}_{-0.12} $ & $ 16.64^{+0.32}_{-0.41} $ &  $ 17.95^{+0.15}_{-0.19} $   \\
6 & 0.743& $ 13.9^{+0.10}_{-0.12} $ &  $ 17.32^{+0.50}_{-1.01} $ & $ 17.76^{+0.23}_{-0.15} $ & $ 4.12^{+0.11}_{-0.12} $ & $ 16.13^{+0.33}_{-0.30} $ &  $ 17.71^{+0.16}_{-0.15} $   \\
7 & 0.867& $ 15.8^{+0.20}_{-0.19} $ &  $ 16.1^{+0.86}_{-3.1} $ & $ 17.85^{+0.15}_{-0.15} $ & $ 3.84^{+0.11}_{-0.10} $   & $ 16.42^{+0.27}_{-0.31} $ &  $ 17.84^{+0.13}_{-0.15} $   \\
8 & 0.990& $ 13.6^{+0.51}_{-0.29} $ &  $ 17.22^{+0.44}_{-4.22} $ & $ 17.78^{+0.14}_{-0.17} $ & $ 3.86^{+0.10}_{-0.11} $ & $ 16.33^{+0.32}_{-0.29} $ &  $ 17.81^{+0.15}_{-0.14} $   \\    
9 & 1.114 & $ 12.9^{+0.08}_{-0.10} $ &  $ <16.64 $ & $ 17.87^{+0.10}_{-0.10} $ & $ 3.89^{+0.07}_{-0.12} $  	& $ 16.60^{+0.32}_{-0.21} $ &  $ 17.93^{+0.15}_{-0.10} $   \\
\enddata
\tablecomments{
$^{1}$We imposed a lower limit of 13 for the \lognh\ of Fe XXV and Fe XXVI column density.}
\end{deluxetable*}

\begin{figure*}
\begin{center}
\includegraphics[width=0.85\textwidth]{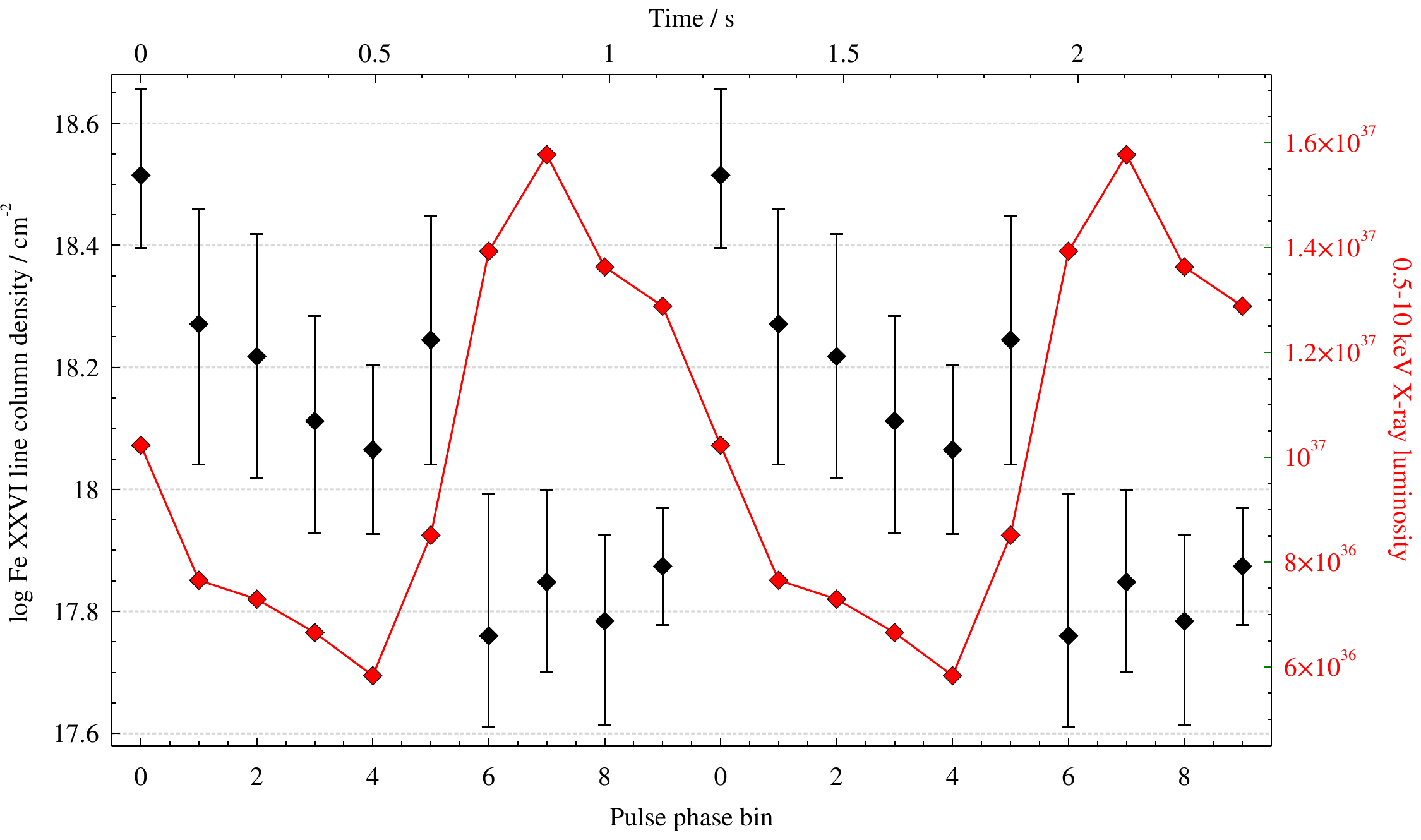}
\caption{Fe XXVI ion column density of the Her X-1 disk wind (black points) versus the pulse phase bin (on the bottom X-axis) and versus time (on the top X-axis). In red color, we show the $0.5-10$ keV unabsorbed Her X-1 luminosity versus the phase bin and time. The Fe XXVI ions respond to the variation in X-ray luminosity over the pulsation period. \label{101_slab}}
\end{center}
\end{figure*}

The constraints on the Fe XXV ion column density are generally quite loose and hard to interpret in terms of X-ray pulsation, as the ion is significantly detected only in a few pulse phase bins. The measurements are listed in Table \ref{pnresults_table}. On the other hand, the Fe XXVI ion column density is well measured in all of the 10 phase bins and it shows a very clear anti-correlation with the $0.5-10$ keV Her X-1 luminosity. The results are shown in Table \ref{pnresults_table} as well as in Fig. \ref{101_slab}, where the column density in the Fe XXVI ion is plotted versus the pulse phase bin number. The same figure also contains the Her X-1 $0.5-10$ keV unabsorbed luminosity for each phase bin (in red). We observe that while the \lognh\ of Fe XXVI column density is around $18.1-18.2$ when the X-ray luminosity is low, it significantly decreases when the X-ray luminosity increases, down to about \lognh$\sim17.8$. In other words, some of the Fe XXVI state atoms are fully ionized when the observed X-ray luminosity rapidly increases and the plasma ionization rises in response (see Fig. \ref{ion_con} in Appendix \ref{app:ion_con}), but recombine again into Fe XXVI when the X-ray luminosity returns to the original value. These results can be directly compared with the \textsc{tpho} simulations from Section \ref{sec:tpho}, shown in Fig. \ref{tpho_results}. The very quick response of the plasma to variations in the X-ray luminosity on 1~s timescales, with no obvious time delays indicates that the plasma density is high, at least $10^{12}$ \pcmq.

\subsection{Physical photoionization disk wind modeling}

As a second approach, we use the photoionization spectral model \textsc{pion} to describe the disk wind features. \textsc{pion} \citep{Miller+15, Mehdipour+16} calculates the transmission through a slab of plasma by self-consistently determining the ionizing balance using the spectral energy distribution (SED) of the currently loaded continuum model. It is therefore one of the highest fidelity, currently available physical photoionization models. Practically, the SED shape and luminosity used by \textsc{pion} will contain some extrapolation of the models we are using to describe the Her X-1 continuum (particularly the hard Comptonization and the soft blackbody) beyond the available $0.5-10$ keV band of EPIC pn. This is one of the drawbacks of the photoionization wind analysis given our dataset with a limited energy range. 

\textsc{pion} assumes photoionization equilibrium of wind plasma for each pulse phase bin, which may not be a correct assumption. However, at these precession phases (0.02-0.04), the wind number density is expected to be reasonably high, and so the plasma should not be too far from the equilibrium. Even if that was not the case, the ionization parameter, which in these fits is determined primarily from the ratio of the Fe XXV and XXVI transitions, would still offer a very good guideline on the immediate ionization state of the wind. As discussed in Section \ref{sec:tpho}, we prefer not to directly fit the observed spectra with the time-dependent photoionization model \textsc{tpho} due to the constant SED approximation in the current version of \textsc{tpho}.

We use the \textsc{pion} model to determine the ionization parameter \logxi\ of the disk wind at each phase bin. The column density, outflow velocity and velocity width parameters are fixed to the best-fitting values from the time-averaged fit of this observation, as they are unlikely to vary in a correlated way over the pulsation period. We also adopt the best-fitting wind elemental abundances from \citet{Kosec+23a}, which are (N/O)$=3.4_{-0.8}^{+0.6}$, (Ne/O)$=2.3_{-0.5}^{+0.4}$ and (Fe/O)$=2.1 \pm 0.3$.

We note that we encountered an issue while fitting pulse phase bin number 8, where a very low \logxi=$3.26_{-0.10}^{+0.11}$ was obtained initially. Our interpretation is that this issue occurred because of an incorrect extrapolation of the Her X-1 SED shape above 10 keV (beyond the EPIC pn energy band). The model originally fitted the (extrapolated) $0.1-100$ keV Her X-1 luminosity to be a factor of 10 higher than that of the two neighbouring phase bins. From hard X-ray pulse-resolved X-ray studies \citep[e.g.][]{Fuerst+13}, we know that such a large X-ray luminosity cannot be correct. Specifically, this is likely an issue with the hard Comptonization model \textsc{comt} - but it is not a technical issue. Apparently, the Her X-1 continuum during phase bin 8 decreases faster beyond 10 keV than predicted by \textsc{comt}. We used an extra exponential cutoff model (\textsc{etau} in \textsc{spex}) to adjust the calculated $0.1-100$ keV luminosity of bin 8 to be the average of the two neighbouring bins (7 and 9). Doing so increased the best-fitting ionization parameter to \logxi=$3.86_{-0.11}^{+0.10}$. We note that none of the remaining pulse phase bins had an abnormal extrapolated $0.1-100$ keV luminosity. Nevertheless, this issue underlines the need for good coverage of the full X-ray band where Her X-1 peaks ($0.3-50$ keV) in future studies. Unfortunately, no simultaneous \nustar\ data are available for \xmm\  observation 865440101.

\begin{figure*}
\begin{center}
\includegraphics[width=0.85\textwidth]{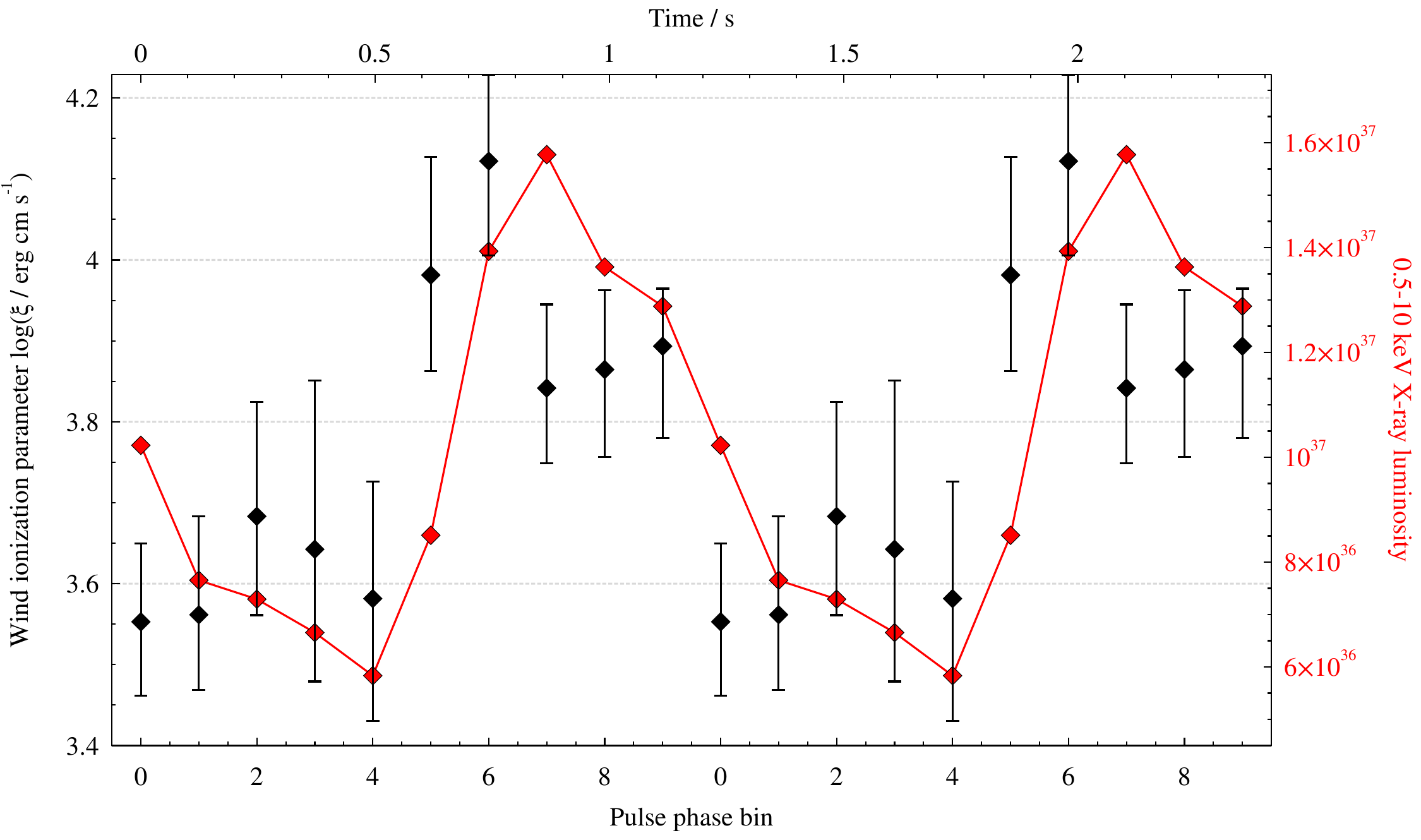}
\caption{Disk wind ionization parameter \logxi\ (black points) versus the pulse phase bin (on the bottom X-axis) and versus time (on the top X-axis). In red color, we show the $0.5-10$ keV unabsorbed Her X-1 luminosity versus the pulse phase bin and time. Significant variation of the ionization parameter is observed over the pulsation period, as the wind ionization state responds to the X-ray variation. \label{101_pion}}
\end{center}
\end{figure*}

The best-fitting ionization parameter \logxi\ is shown in Fig. \ref{101_pion} versus the pulse phase bin number, alongside the $0.5-10$ keV luminosity. For comparison with the phenomenological fitting approach, we used the \textsc{ascdump} command in \textsc{spex} to determine the column densities of the Fe XXV and XXVI ions in the \textsc{pion} model. All these measurements are given in Table \ref{pnresults_table}.

A response of the ionized wind to the X-ray pulsation is clear - the pulse phase bins with higher $0.5-10$ keV luminosity show consistently higher ionization parameters. Again, comparing with the \textsc{tpho} simulations, we conclude that the wind number density must be at least $10^{12}$ \pcmq\ for such a quick response, confirming our time-averaged ionization state modeling results. The Fe XXVI column density is anti-correlated with the ionization parameter (as it should be, see Fig. \ref{ion_con}) and in good agreement with the phenomenological wind modeling for most phase bins. The exception is phase bin 5, where the $0.5-10$ keV luminosity is beginning to rise slightly. The Fe XXVI column density still remains relatively high, but in the photoionization analysis, the ionization parameter has already significantly increased. We could not find the cause of this discrepancy, but it could again be related to the underlying \textsc{pion} modeling of the SED from the extrapolated Her X-1 continuum model.

\section{\xrism\ simulations} \label{sec:xrism}

In the previous section, we showed that the highest quality \xmm\ observation of the disk wind in Her X-1 allows us to directly detect the variation of its ionization properties over the pulsation period. Lower quality wind observations (with shorter exposure times) at low Her X-1 precession phases, when the wind signatures are strong, may also be sufficient to detect this variation (at lower significance). However, the wind absorption significantly weakens with increasing precession phase \citep{Kosec+23a} and so its variation over the pulsation period is unlikely to be detectable with current X-ray instruments such as EPIC pn onboard \xmm, \nicer\ or \chandra\ at higher precession phases.

In September 2023, a new high-resolution X-ray spectroscopy mission \xrism\ \citep{XRISM+20} was launched. \xrism\ offers excellent spectral resolution of about 5 eV across the $0.5-10$ keV band thanks to its Resolve micro-calorimeter instrument. \xrism\ is therefore perfect for spectroscopic observations of the Fe K band, which in Her X-1 contains the important Fe XXV and XXVI wind absorption lines. These transitions are crucial for the determination of the outflow ionization state. Secondly, \xrism\ offers a very good effective area for a high-resolution instrument. Fig. 6 of \citet{Kosec+23c} shows simulated spectrum from a 10ks exposure \xrism\ observation of Her X-1 in the Short High state, focusing on the Fe K band. The data quality of this very brief exposure simulation, in the Short High state (which is a factor of $2-3$ fainter than the Main High state) is excellent and the disk wind absorption lines are easily spectrally resolved. Even shorter exposures will be sufficient to accurately measure the ionization properties of the outflow of Her X-1. Finally, the Resolve micro-calorimeter offers very good time resolution of less than 1 ms.

We perform \xrism\ simulations in order to determine the performance of this instrument and to understand if its future observations will allow us to directly measure the plasma number density in the outflow of Her X-1, for which we can currently only obtain lower limits. We used publicly available \xrism\ simulation responses\footnote{https://heasarc.gsfc.nasa.gov/docs/xrism/proposals/} and assume that the goal 5 eV spectral resolution will be achieved. We also note that we assumed the gate valve open configuration of the Resolve instrument. Since the disk wind lines which are the most important for the ionization parameter measurement, Fe XXV and XXVI, are located in the Fe K band, a closed gate valve does not mean that \xrism\ cannot perform this science. However, the exposure needs to be increased accordingly (by about 60\%) to reach the quality of results in these simulations.

Same as in Section \ref{sec:tpho}, we perform spectral simulations using the time-dependent photoionization model \textsc{tpho}. As the X-ray pulsation lightcurve, we again choose the $0.5-10$ keV unabsorbed lightcurve from observation 0865440101. The underlying emission continuum is the best-fitting Her X-1 continuum from that observation, but scaled by the $0.5-10$ keV lightcurve in each of the 10 pulse phase bins (each of the individual spectral components is scaled by the same factor). This way we ensure a constant shape of the continuum SED as required by the current version of \textsc{tpho}.

We probe the detectability of the wind response to X-ray pulsation with \xrism\ for different precession phases of Her X-1. Therefore, we test a number of different wind properties (column densities, ionization parameters). We use the representative wind properties determined in Appendix \ref{app:meanwind} for precession phases of 0.03, 0.06, 0.09 and 0.13. Finally, we test a broad range of possible wind number densities, following Section \ref{sec:tpho}: from very low ($10^{10}$ \pcmq) to high ($10^{14}$ \pcmq) densities.

In the real observations of Her X-1 with \xrism\ throughout the precession phase, the underlying time-averaged as well as pulse-resolved emission continuum of the source is bound to evolve. However, this is unlikely to introduce significant extra uncertainties on the recovered wind properties as the overall time-averaged continuum flux does not significantly change throughout the Main High state and the wind absorption lines are narrow and easily separable from the spectral continuum at the excellent \xrism\ resolution.

Following the approach discussed in Section \ref{sec:tpho}, we introduce the \textsc{tpho} wind absorption component, and evolve the plasma for 5 full pulsation cycles to obtain a stable ionization response. Afterwards, we simulate \xrism\ spectra for 10 pulse phase bins over the rotation period of Her X-1. We mimic a 25 ks total exposure \xrism\ observation by simulating 10 2.5 ks exposure phase bins. The mock \xrism\ spectra are saved individually, and each of them is fitted using the \textsc{pion} photoionization model to determine the best-fitting immediate ionization parameter of the wind plasma. This procedure is followed for all simulated wind number densities, and for \xrism\ simulations at all the different Her X-1 precession phases.

\begin{figure*}
\begin{center}
\includegraphics[width=\textwidth]{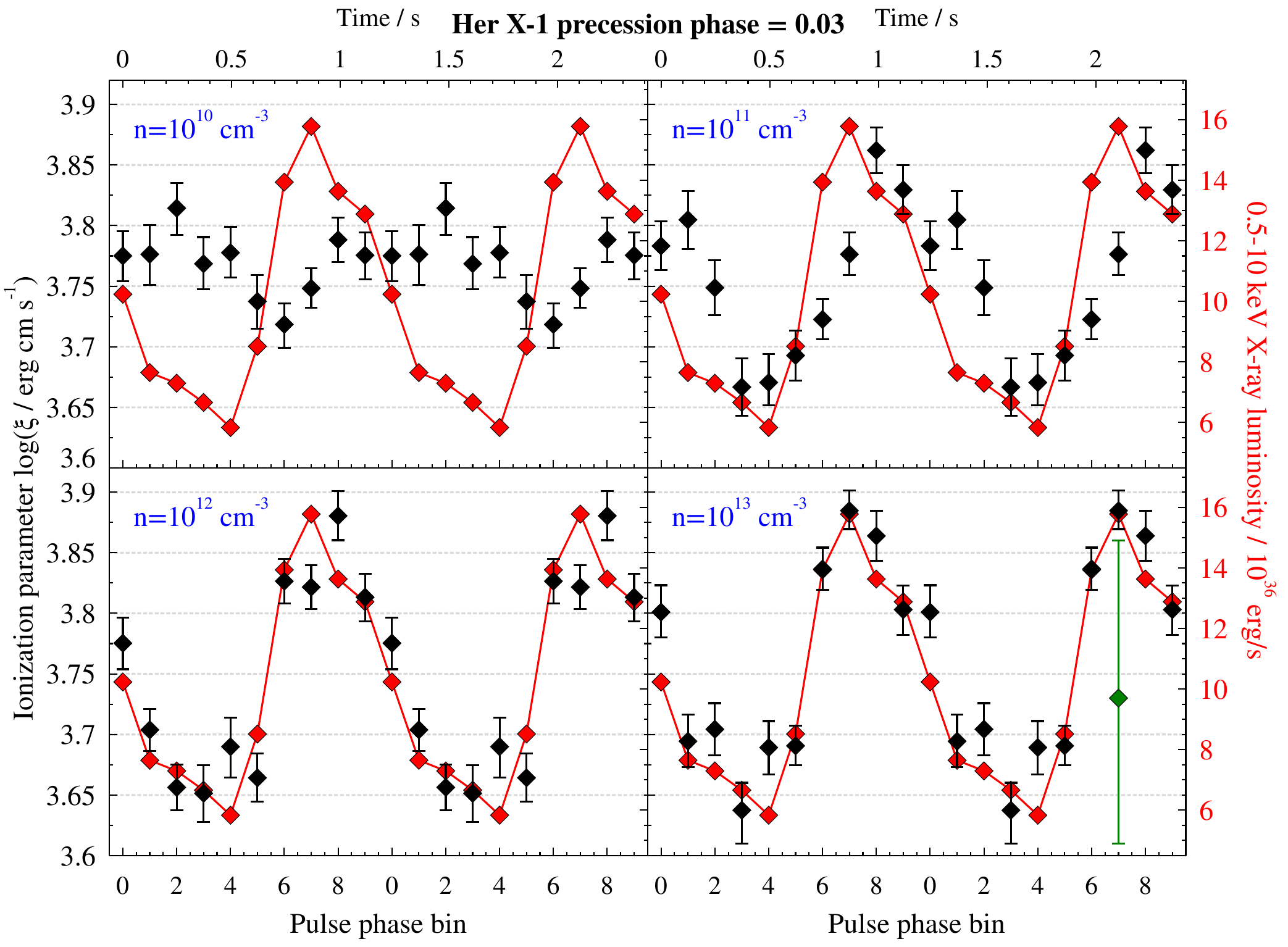}
\caption{Pulse-resolved analysis (10 bins per pulsation period) of simulated 25 ks \xrism\ observations of Her X-1 at precession phase 0.03 (column density $10^{23}$ \pcmq, \logxi\ of 3.77). The different panels show simulations with different input plasma number densities. The red diamonds show the underlying X-ray pulsation lightcurve, and the black diamonds show the best-fitting ionization parameter \logxi\ for each simulated \xrism\ pulse phase bin. For comparison, the green data point in the bottom right panel shows the mean uncertainty on \logxi\ measurement from the analysis of \xmm\ observation 0865440101 (Section \ref{sec:results}).  \label{xrism0p03}}
\end{center}
\end{figure*}

The results for the simulated \xrism\ observation at precession phase 0.03 are shown in Fig. \ref{xrism0p03}. The figure is composed of 4 panels corresponding to different input plasma number densities ($10^{10}$ through $10^{13}$ \pcmq). Each panel is similar to Fig. \ref{101_pion} but instead of EPIC pn data, it contains a 25 ks \xrism\ observation simulation. We conclude that \xrism\ can easily and significantly detect any plasma ionization variations in the simulated exposure. For very low number densities ($10^{10}$ \pcmq), the outflow response is very weak, as expected. At higher densities ($10^{11}$ \pcmq), the response is stronger but very delayed. At high densities ($10^{12}$ and $10^{13}$ \pcmq), the response is fast and strong. We also simulated \xrism\ spectra for plasma number density of $10^{14}$ \pcmq\ but did not find significant visual differences between the results from those at the number density of $10^{13}$ \pcmq. Even though no differences may be seen visually, direct fitting with a future version of the \textsc{tpho} model (incorporating a variable SED) will likely place strong constraints on the plasma number density. Even in the worst case, it will result in a significantly more constraining number density lower limit than the \xmm\ observations.

As the \xrism\ data have excellent spectral resolution and the wind absorption is strong at precession phase 0.03, the uncertainties on simulated data points are very small. The mean uncertainties on the ionization parameter are about $\Delta$\logxi$\sim0.02$, whereas in the \xmm\ \textsc{pion} analysis (Section \ref{sec:results}) they were on average $\Delta$\logxi$\sim0.12$. To show this visually, we added the mean \xmm\ \logxi\ uncertainty in the bottom right panel of Fig. \ref{xrism0p03}.

\begin{figure*}
\begin{center}
\includegraphics[width=\textwidth]{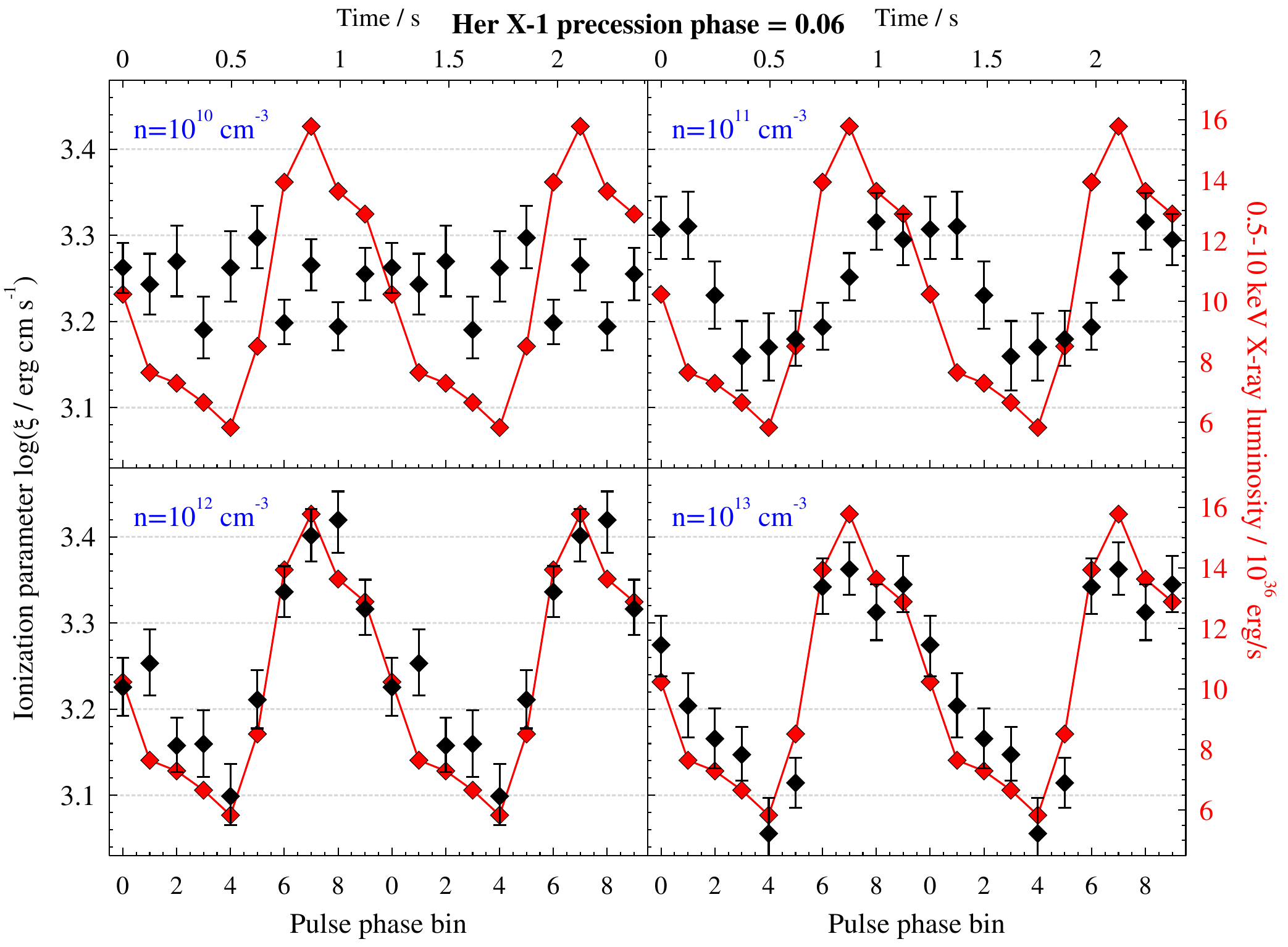}
\caption{Same as Fig. \ref{xrism0p03}, but for representative disk wind properties of Her X-1 at the precession phase of 0.06 (column density $10^{22}$ \pcmq, \logxi\ of 3.25). \label{xrism0p06}}
\end{center}
\end{figure*}

\begin{figure*}
\begin{center}
\includegraphics[width=\textwidth]{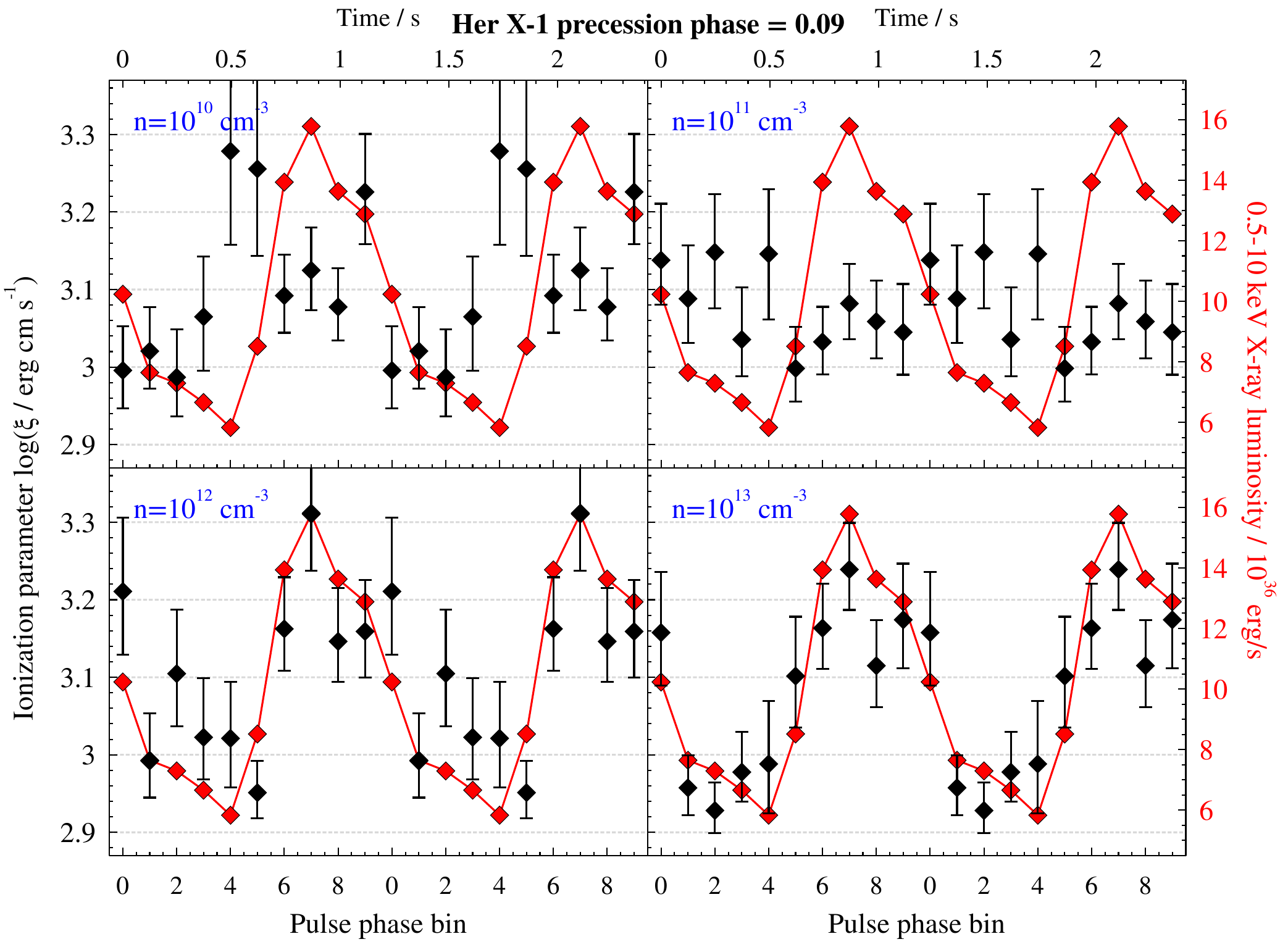}
\caption{Same as Fig. \ref{xrism0p03}, but for representative disk wind properties of Her X-1 at the precession phase of 0.09 (column density $3\times10^{21}$ \pcmq, \logxi\ of 3.1). \label{xrism0p09}}
\end{center}
\end{figure*}

We also performed \xrism\ simulations for wind properties at precession phases of 0.06 and 0.09. The results are shown in Fig. \ref{xrism0p06} and \ref{xrism0p09}. The wind absorption is weaker in comparison with phase $\sim0.03$ since the column density is lower, but the response of plasma to X-ray pulsation is still detectable in a 25 ks exposure.

Finally, we performed a 25 ks simulation for wind properties equivalent to precession phase of 0.13. Here the wind absorption is so weak that the response is no longer detectable with \xrism\ using 10 pulse phase bins. The outflow is still significantly detected in the time-averaged 25ks spectrum, and so a coarser pulse-resolved analysis (e.g. with 5 bins per pulse period) may still be capable of detecting the wind response provided the number density is sufficient.

\section{Discussion} \label{sec:discussion}

We study the ionization response of the accretion disk wind in Hercules X-1 to the X-ray pulsations at the neutron star rotation frequency. We find that given the number density limits derived from previous wind spectroscopic observations, the response can be sufficiently fast to be detected in X-ray observations. We then perform pulse-resolved analysis of the best-quality \xmm\ observation of the Her X-1 disk wind and detect this ionization response.

We use two different spectral modeling approaches to measure how the ionization state of the wind responds to the X-ray pulsations in the best-quality Her X-1 disk wind observation (0865440101). In the first method, we track the pulse-resolved evolution of individual ion column densities (Fe XXV and XXVI). Secondly, we track the ionization parameter \logxi\ of disk wind plasma using the physical photoionization model \textsc{pion}. The phenomenological analysis shows that the Fe XXVI column density sharply drops just as the X-ray flux increases (and some of the Fe XXVI ions are fully ionized), and returns back to the high value when the X-ray flux reduces. The Fe XXV ion is detected only weakly throughout the pulse period, but its column density evolution agrees with the Fe XXVI results. The ionization parameter evolution in the \textsc{pion} physical analysis is broadly consistent with the phenomenological results, despite one inconsistency which may be due to the measurement of the underlying X-ray illumination SED (which must be extrapolated from the available $0.5-10$ keV energy band). At these high ionization levels, the \textsc{pion} model determines the \logxi\ parameter primarily from the strengths of the Fe XXV and XXVI lines. We observe good agreement between the column densities of these Fe ions as predicted by \textsc{pion} and measured by the \textsc{slab} model (Table \ref{pnresults_table}).

Unfortunately, during \xmm\ observation 0865440101 there was no simultaneous observation with the \nustar\ X-ray telescope, which would have been able to capture the hard X-ray SED of Her X-1 up to 79 keV. Additionally, we are unable to use the simultaneous RGS grating spectra for the pulse-resolved analysis as the instrument was operated in normal Spectroscopy mode during the observation, which has a frame time much longer than the 1.24~s X-ray pulsation period. With RGS data, we would have been able to significantly decrease the uncertainties on the \logxi\ measurements as well as measure column densities in other strong disk wind ions such as O VIII and Ne X. 

Both the phenomenological and physical modeling methods indicate a fast response of the disk wind to X-ray pulsations, and we do not detect any clear delay in this response. Comparing this with the \textsc{tpho} simulations (Fig. \ref{tpho_results}), the EPIC pn results suggest that the wind number density must be significantly higher than $10^{11}$ \pcmq, most likely at least $10^{12}$ \pcmq. This is a completely independent confirmation of the lower density limit obtained from the time-averaged wind spectroscopic analysis, which indicates that $n\gtrsim4\times10^{12}$~\pcmq.

The best-quality \xmm\ EPIC pn dataset shows variations in the ionization state of the wind at high statistical significance despite the complexity of the Fe K region in Her X-1, thanks to its long exposure and the strength of the ionized absorption from the wind at this precession phase. At higher precession phases when the wind absorption is weaker, the effect is likely undetectable with \xmm. 


\chandra\ HETG gratings \citep{Canizares+05} offer better spectral resolution than EPIC pn in the Fe K band, but also a poorer effective area, requiring more exposure time. In the Continuous-Clocking (CC) mode (3 ms timing resolution), they are capable of detecting the wind time variation over the pulsation period, but again only when the wind absorption is strong. HETG spectra in the CC-mode were previously used to perform a phase-resolved analysis of disk wind absorption during the heartbeat state of the black hole X-ray binary GRS 1915+105 \citep{Neilsen+11}. The heartbeats of GRS 1915 are quasi-periodic with a typical duration of about 50 s. With the phase-resolved analysis, \citet{Neilsen+11} detected variations in the Fe XXV/Fe XXVI absorption line properties on timescales as short as 5 s.

We conclude that with the current X-ray instruments, we can set a lower limit on the wind number density at low precession phases (low heights above the accretion disk), but cannot directly measure its value. At greater heights above the disk, we are likely unable to place any relevant constraints on the wind number density via time-dependent photoionization modeling. However, in order to significantly constrain the wind mass outflow rate, energetics and launching mechanism, we ideally need multiple number density measurements (or strong constraints) at different locations within the outflow.

In Section \ref{sec:xrism}, we find that the best instrument to pursue an accurate wind number density measurement throughout the outflow is the recently launched \xrism\ telescope, which combines excellent spectral resolution with a very good effective area in the Fe K band. Our \xrism\ simulations of the Her X-1 disk wind response to X-ray pulsation show that this effect is detectable throughout a significant fraction of the Main High state of Her X-1 (at least up to and including the precession phase of 0.09). At very low phases, the wind number density must be high (n~$\gtrsim4\times10^{12}$ \pcmq) and so the wind response is very fast. Even with \xrism, it may only be possible to obtain a lower limit to the wind density. In any case, thanks to the superior \xrism\ data quality, the limit is likely going to be more informative than the one from \xmm\ data. At higher precession phases that sample greater heights along the wind streamlines, unless the wind clumps up very quickly, the number density will most likely decrease. Therefore, the wind response to pulsations should become progressively slower. If the density drops below about $10^{12}$ \pcmq, this change will be directly (and accurately) measurable with \xrism. At even higher phases, the density may drop to such low values that the wind would no longer be capable of any response to the X-ray pulsations.

To sum up, \xrism\ observations at low precession phases will put at least a lower limit on the wind density, while observations at greater phases will most likely be capable of directly measuring the density value. Finally, observations at the highest phases may even put upper limits on the wind density based on the lack of any response, while the photoionization balance analysis will still place a lower limit on the density. Therefore, a coordinated observational campaign on Her X-1 with \xrism, sampling its disk wind properties throughout a significant fraction of the Main High state will allow us to measure the number density of the wind at multiple locations and heights above the disk. This accurate measurement of the number density and the wind properties (in 2D) will constrain the total mass outflow rate and energetics, enabling us to infer the outflow launching mechanism and the impact it has on the X-ray binary system and its surroundings.

Our analysis of the $0.5-10$ keV \xmm\ spectra highlights the need to accurately constrain the pulse-resolved SED of Her X-1 to model the disk wind variation correctly. \xrism, with a useful energy range from 0.3 keV of up to about 17 keV, will be able to place better constraints on this SED shape (which peaks around $20-25$ keV) than \xmm. However, at least partial simultaneous coverage of any \xrism\ observations with a hard X-ray observatory such as \nustar, would still be greatly beneficial for a more robust SED determination. This would extend the visibility up to 80 keV and cover practically all of the SED of Her X-1.


\subsection{Other potential applications of time-dependent photoionization analysis in X-ray binaries}

As shown in this study, time-dependent photoionization analysis can be a powerful tool to constrain the properties of ionized plasma and outflows in X-ray binaries exhibiting significant X-ray time variability. In many cases, the variability of these systems may be too fast to achieve sufficient signal-to-noise for such a spectral analysis. However, the repetition of X-ray flux variations provides the crucial boost to signal-to-noise in our case of Her X-1. Such a strategy may be applied to other systems. X-ray bursts are an excellent example of strong and fast X-ray variability, but as they are very brief, stacking of many bursts may be required to detect the ionization response of any plasma signatures in their spectra \citep{Strohmayer+19}.

Unfortunately, accretion disk winds do not appear to be common among the population of accreting X-ray pulsars which are the best candidates for this kind of analysis as they often pulsate with a period of about 1~s. However, time-dependent photoionization modeling could potentially be leveraged to study the properties of donor star outflows in high-mass X-ray binaries such as Vela X-1 \citep[e.g.][]{Kretschmar+21}. Assuming that the pulsation period is either much lower or much higher than the donor wind clump-crossing time and the plasma number density is sufficiently high, the ionization response should be detectable in a high-quality X-ray dataset.

Another type of objects pulsating at similar periods ($\sim$1~s) are the neutron star powered ultraluminous X-ray sources \citep[ULXs, e.g.][]{Bachetti+14}. A major issue is that they are much fainter than Galactic X-ray binaries due to their Mpc distances. Secondly, their pulsed fraction (giving the strength of the periodic variability) is in most cases lower than the $40-50$\% fraction observed in Her X-1, and only a few have shown evidence of ionized outflows to date \citep{Kosec+18b, vandenEijnden+19}. An exception is NGC 300 ULX-1, a pulsating ULX which showed both an ionized outflow \citep{Kosec+18b} as well as high pulsed fraction \citep[of more than 50\%,][]{Carpano+18}. Observations of similar sources with future high-resolution, high-effective area X-ray telescopes such as \athena\ (yielding very high quality spectra) would be an excellent opportunity for time-dependent photoionization studies, allowing us to improve our understanding of ULX outflows \citep{Pinto+23}.

The final class of objects we consider here are the accretion powered X-ray millisecond pulsars \citep[AMXPs,][]{Wijnands+04, DiSalvo+22}. Their X-ray pulsations could be leveraged in a time-dependent photoionization modeling (assuming thath the pulsed fraction is sufficient), and some of them show evidence for ionized outflows \citep{Marino+22}. However, their X-ray variability timescales, given the rotation periods of 100 Hz and above may be too fast. Unless the plasma has a very high number density, it will not have sufficient time to respond to these periodic variations.

In the present study, we exclusively measured the ionization response of a disk wind through its absorption lines. In principle, a similar study can also be performed for any ionized emission lines in a time variable X-ray binary spectrum. Such analysis is more complicated as the ionized emission can originate in a much more spatially complex and extended region in comparison with the ionized absorption (which necessarily originates somewhere along our line sight towards the X-ray source). Additionally, it will be important to consider how the light travel time effects (which do not play a role in absorption) influence the observed line response. Nevertheless, the response of emission lines observed in accreting pulsars to X-ray pulsation could still reveal valuable information about their origin. 

In the most ideal accretion disk wind study, we would combine the time-variable response of both the absorption and the emission from the wind, as well as use the Doppler information in the wind spectral lines. As a result, we would be able to obtain an equivalent of a Doppler tomograph \citep{Marsh+88, Marsh+05}, but sampled by the illumination pattern of the X-ray pulsar rather than by the orbital motion of the binary. In order to achieve this, excellent spectral resolution is required and both the absorption and the emission lines must be detected at very high statistical significance. Such analysis may be possible with \xrism\ or \athena\ \citep{Nandra+13} observations of Her X-1.

\section{Conclusions} \label{sec:conclusions}

We performed an extensive pilot study assessing how the ionization state of the disk wind in Hercules X-1 responds to the periodic pulsation of the illuminating X-ray continuum, which is introduced by the neutron star rotation. We specifically studied how this plasma state response can be used to uniquely measure the wind number density. Our findings can be summarized as follows:

\begin{itemize}
    \item From previous time-averaged X-ray spectroscopic observations of Her X-1, we were able to constrain the wind number density to be between $10^{9}$ and $10^{14}$ \pcmq\ throughout the Main High state of Her X-1. At low precession phases (low disk wind heights above the accretion disk), the number density must be on the high end of this range ($>10^{12}$ \pcmq), but it can decrease to lower values of the range at higher precession phases (and greater heights above the disk).
 
    \item Based on our simulations using the time-dependent photoionization model \textsc{tpho}, at low precession phases (where the wind number density is high), the ionization response of the wind to X-ray pulsation of Her X-1 should be fast and strong, and directly detectable. At higher precession phases, the wind density can be lower, possibly resulting in slower and weaker ionization response and, for the lowest allowed number densities, no response at all.

    \item By performing a pulse-resolved analysis of the best-quality \xmm\ observation of the Her X-1 disk wind, taken at a low precession phase, we directly detected this wind response to X-ray pulsation. Variations over the pulsation period are observed in both the Fe XXVI ion column density (using phenomenological spectral modeling) and in the wind ionization parameter (using physical photoionization modeling) with no detectable time delays, indicating a wind number density of at least $10^{12}$~\pcmq.

    \item We showed that the recently launched \xrism\ X-ray telescope will be capable of detecting the Her X-1 wind variations over the pulsation period throughout a significant fraction of the Main High state. Finally, we discussed how a coordinated observational campaign with \xrism\ would allow us to accurately measure the wind number density at different points within the outflow, constraining the wind mass outflow rate, its energetics and the launching mechanism.
\end{itemize}


\begin{acknowledgments}
 Support for this work was provided by the National Aeronautics and Space Administration through the Smithsonian Astrophysical Observatory (SAO) contract SV3-73016 to MIT for Support of the Chandra X-Ray Center and Science Instruments. PK and EK acknowledge support from NASA grants 80NSSC21K0872 and DD0-21125X. Support for this work was provided by NASA through the NASA Hubble Fellowship grant HST-HF2-51534.001-A awarded by the Space Telescope Science Institute, which is operated by the Association of Universities for Research in Astronomy, Incorporated, under NASA contract NAS5-26555. CP acknowledges support by INAF Large Grant BLOSSOM and PRIN MUR 2022 SEAWIND.
\end{acknowledgments}

%

\vspace{5mm}
\facilities{\xmm, \xrism
}


\software{SPEX \citep{Kaastra+96}, Veusz, Stringray \citep{Huppenkothen+19, Stingray+19}
          }




\appendix

\section{Representative disk wind properties of Her X-1 during the Main High state}
\label{app:meanwind}

In order to determine meaningful plasma number density range for the disk wind of Her X-1 (in Section \ref{sec:density_constraints}), we estimate representative outflow properties at different Her X-1 precession phases using previous X-ray spectroscopic observations. The measurements at different phases effectively correspond to wind properties sampled at different heights above the disk.

\begin{figure}
\begin{center}  
\includegraphics[width=0.5\columnwidth]{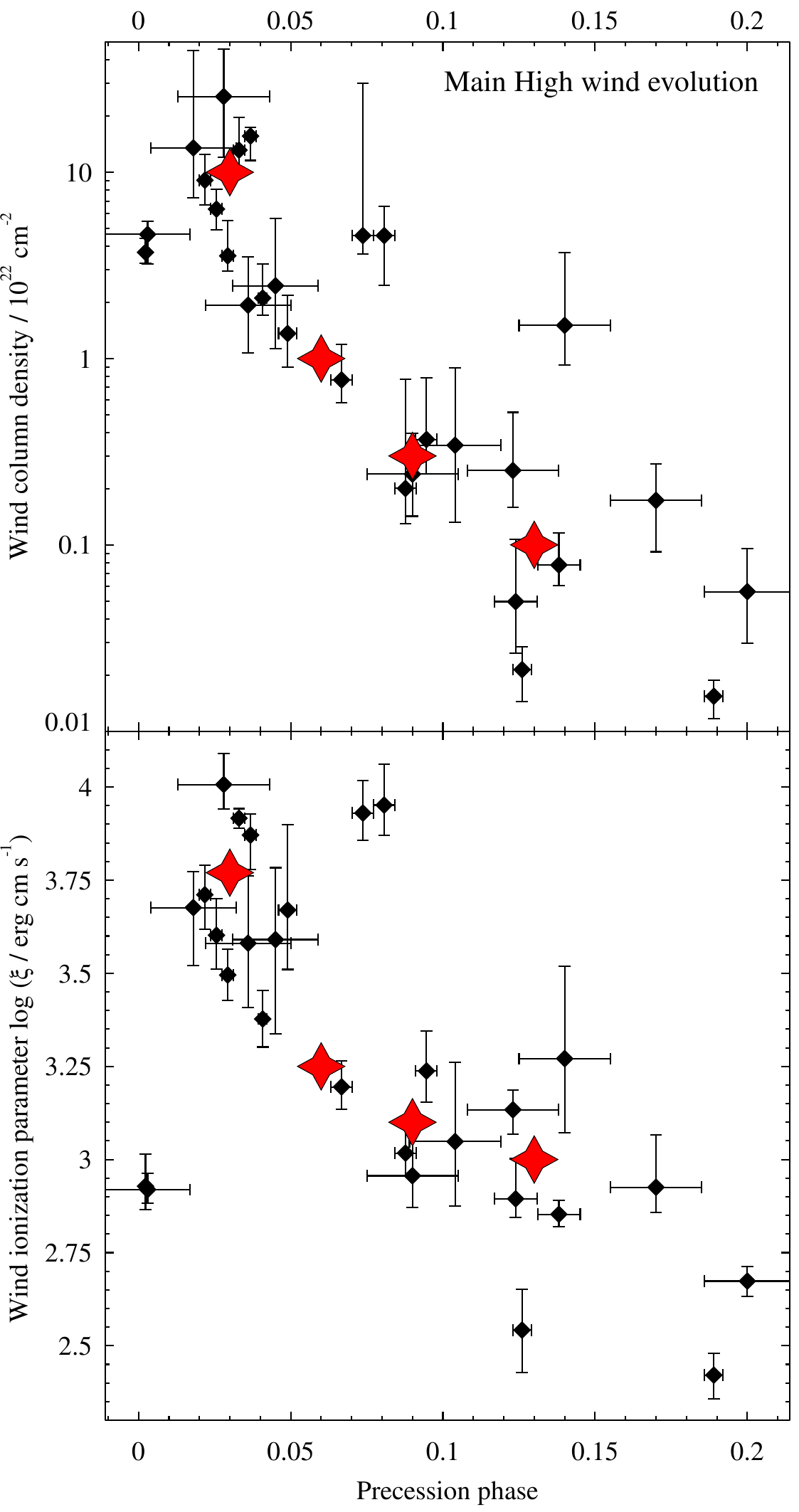}
\caption{Her X-1 disk wind properties versus precession phase. The top panel shows column density, while the lower panel shows ionization parameter \logxi. The black points are the wind measurements from \citet{Kosec+23a} using individual \xmm\ and \chandra\ observations. The red stars are estimated representative wind properties used to determine possible wind density range throughout the Main High state of Her X-1. \label{nh_xi_mean}}
\end{center}
\end{figure}

We use the results of \citet{Kosec+23a}, where we analyzed 28 \xmm\ and \chandra\ observations or observation segments to sample the wind properties throughout the Main High state of Her X-1. The wind column density and ionization parameters are shown in black in Fig. \ref{nh_xi_mean}. The representative wind properties are shown in the same figure using red stars. At precession phase=0.03, these are determined from the highest data quality observation 0865440101, which covers exactly this phase and was analysed in the Methods section of \citet{Kosec+23a}. The column density and \logxi\ values for this phase are direct spectral fit results. At precession phases of 0.06, 0.09 and 0.13, we estimate the representative wind properties by eye from the individual measurements at comparable precession phases. We note that we ignore the two abnormally high column density and \logxi\ measurements (corresponding to two consecutive observation segments around phase $0.07-0.08$) as we consider these datasets to be mis-fitted (it is unlikely that the wind would change properties so rapidly). These representative wind properties are only used to initially estimate the wind number density limits in Section \ref{sec:density_constraints} and as such they do not have to be precise. The wind properties are summarized in Table \ref{meanwind_table}.

\begin{deluxetable}{cccc}
\tablecaption{Representative accretion disk wind properties of Her X-1 at different precession phases.\label{meanwind_table}}
\tablewidth{0pt}
\tablehead{
\colhead{Precession Phase} & \colhead{Column Density} & \colhead{Ionization Parameter} & \colhead{Isotropic Mass Outflow Rate}  \\
\colhead{} & \colhead{$10^{22}$ \pcm} & \colhead{\logxi} & \colhead{$10^{-8}$ $M_{\odot}$ year$^{-1}$} 
}
\startdata 
0.03&10&3.77&5\\
0.06&1&3.25&9\\
0.09&0.3&3.10&25\\
0.13&0.1&3.00&50\\
\enddata
\end{deluxetable}

\begin{figure}
\begin{center}
\includegraphics[width=0.5\columnwidth]{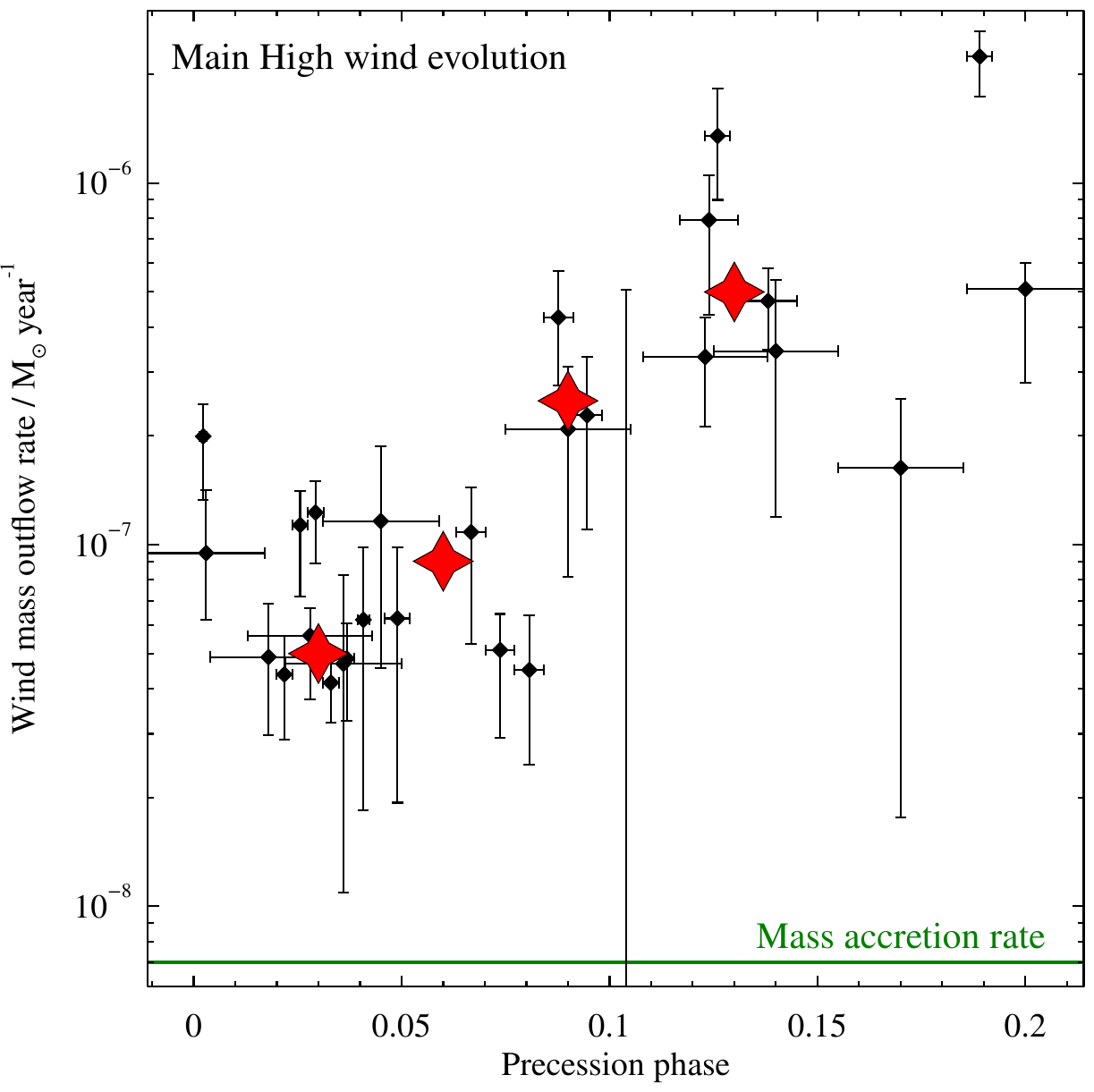}
\caption{Isotropic ($4\pi$) mass outflow rates of Her X-1 versus disk precession phase. Black points are mass outflow rate estimates from individual \chandra\ and \xmm\ observations \citep{Kosec+23a}. Red stars are representative mass outflow rates used to obtain wind number density limits. The green horizontal line is the average measurement of mass accretion rate through the outer disk of Her X-1 \citep{Boroson+07}, indicating that the solid launch angle of the disk wind must be much smaller than $4\pi$. \label{mout_mean}}
\end{center}
\end{figure}

We also calculate representative isotropic mass outflow rates at these precession phases using Equation \ref{eqfinalMout}. The isotropic outflow rate is calculated by assuming full volume filling factor ($C_{\rm{V}}=1$) and full wind launch solid angle ($\Omega=4\pi$). In Fig. \ref{mout_mean}, we show the isotropic mass outflow rates for individual Her X-1 Main High observations from \citet{Kosec+23a}. With red stars, we show the estimated representative mass outflow rates. Again, the first point (at precession phase of 0.03) is calculated from the best-fitting wind properties during observation 0865440101. The remaining 3 points are calculated from the mean wind properties estimated above.

\section{EPIC pn continuum spectral variability over the pulsation cycle}
\label{app:PNspec}

\begin{figure*}
\begin{center}
\includegraphics[width=0.9\textwidth]{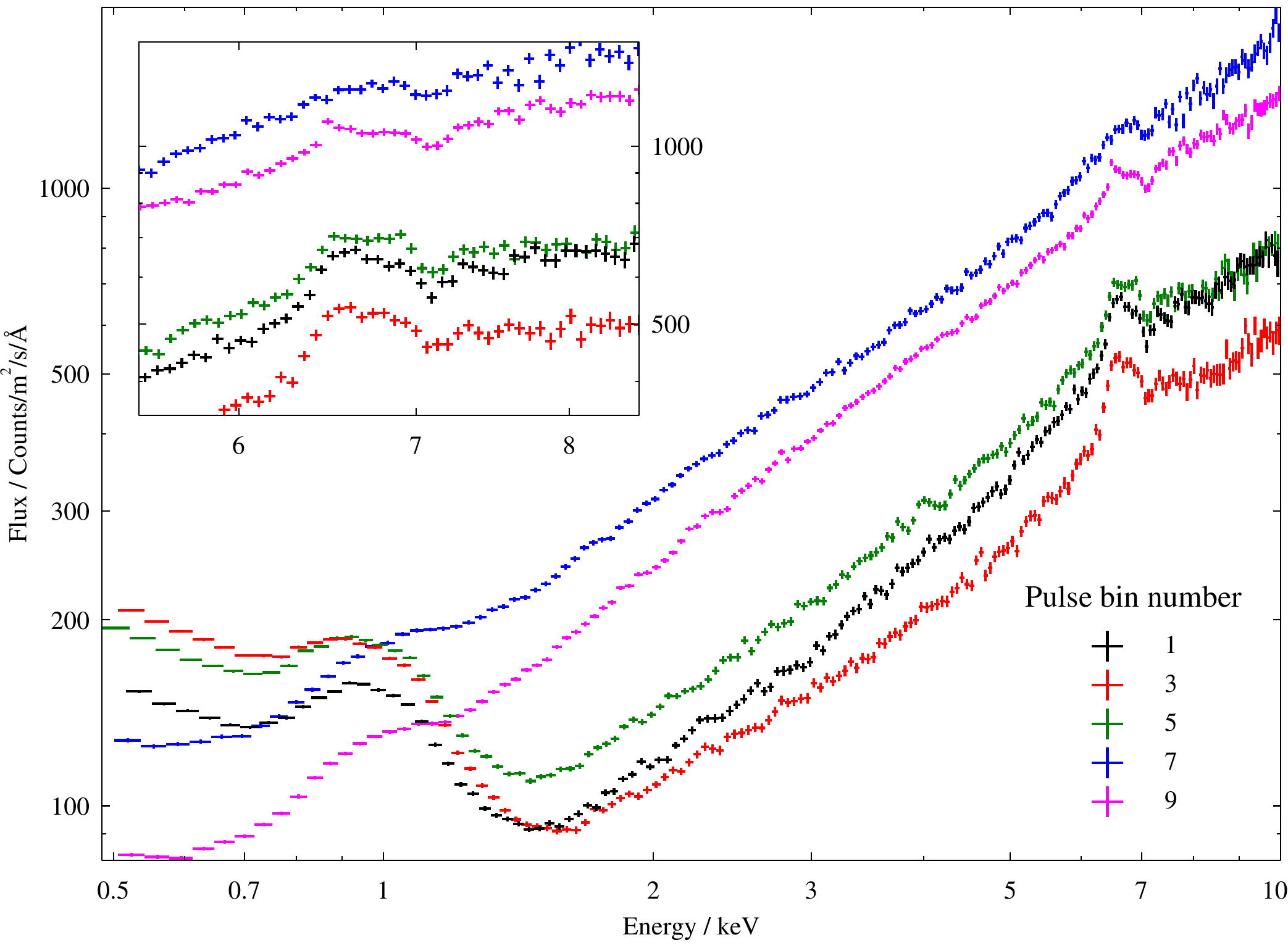}
\caption{EPIC pn spectra from odd-numbered pulse phase bins. Even-numbered phase bins are omitted for visual clarity. The figure shows the complex spectral variation over the pulse period, where the soft X-rays ($<2$ keV) peak at a different pulse phase than the hard X-rays ($>2$ keV). The inset on the top left focuses on the Fe K energy band. \label{PNspecfig}}
\end{center}
\end{figure*}

In Section \ref{sec:data} we split the 1.24~s X-ray pulsation cycle of Her X-1 into 10 phase bins and extracted their EPIC pn spectra. To illustrate the strong X-ray variability over the pulse period, we show the spectra of the odd-numbered phase bins in Fig. \ref{PNspecfig}.

\section{Relative concentrations of the ionic states of Fe}
\label{app:ion_con}

\begin{figure*}
\begin{center}
\includegraphics[width=0.7\textwidth]{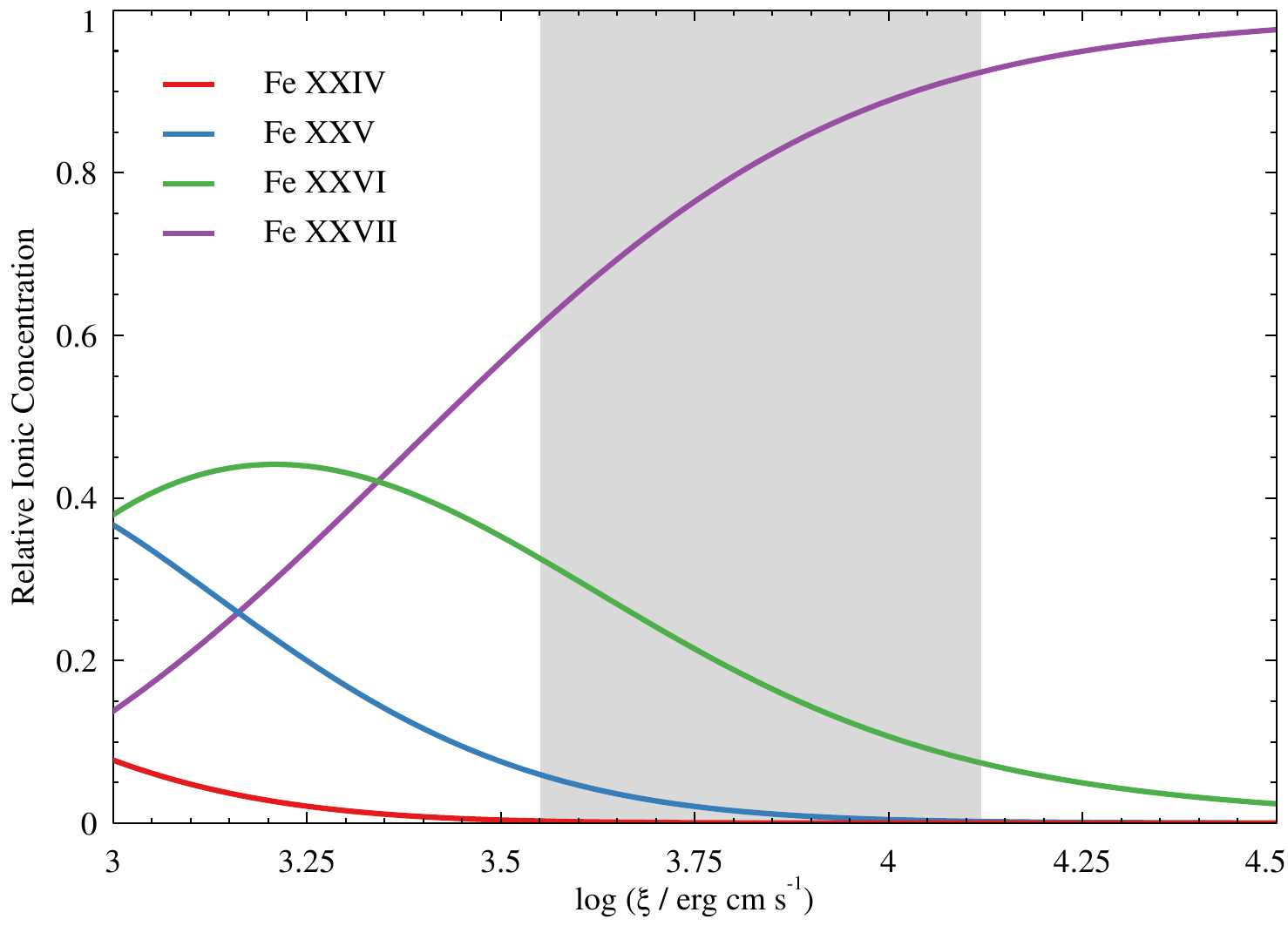}
\caption{Relative concentrations of the 4 highest charged Fe ions - Fe XXIV, XXV, XXVI and XXVII, versus the ionization parameter \logxi. The shaded area indicates the range of the best-fitting ionization parameters from the \textsc{pion} pulse-resolved analysis of \xmm\ observation 0865440101. \label{ion_con}}
\end{center}
\end{figure*}

To understand the variation of the different ionic column densities of Fe over the X-ray pulsation period and how it relates to the ionization parameter from the \textsc{pion} spectral modeling, we calculate the relative ionic concentrations of the highest charged Fe states. We apply the \textsc{Ascdump} command in \textsc{spex} to determine the relative concentrations of Fe XXIV, XXV, XXVI and XXVII for a range of \logxi. We use the time-averaged \xmm\ observation 0865440101 (high flux period only) as the input SED for these calculations. The results are shown in Fig. \ref{ion_con}. The gray shaded region indicates the range of best-fitting \logxi\ from the \textsc{pion} pulse-resolved analysis of observation 0865440101. The individual pulse phase bins have slightly different SEDs, and so their relative concentrations of Fe ions will slightly differ from the values in Fig. \ref{ion_con}. Nevertheless, the figure gives us helpful insights on how the different Fe ions change in concentration with respect to each other as the X-ray flux varies over the pulsation period of Her X-1.

\bibliography{References}{}
\bibliographystyle{aasjournal}



\end{document}